\documentclass[journal,twoside,web]{ieeecolor}
\usepackage{20190813_TCST_rDMM}
\usepackage{cite}
\usepackage{amsmath,amssymb,amsfonts}
\usepackage{algorithmic}
\usepackage{graphicx}
\usepackage{textcomp}

\newcommand{\ihat}{\boldsymbol{\hat{\i}}}
\newcommand{\jhat}{\boldsymbol{\hat{\j}}}

\newcommand{\R}{\mathbb{R}}

\newtheorem{assumption}{\bf Assumption}

\usepackage{algorithm}
\usepackage{algorithmic}

\usepackage{booktabs}
\usepackage{caption}
\usepackage{tabularx}

\newcolumntype{P}{>{\centering\arraybackslash}m{0.3in}}
\newcolumntype{I}{>{\centering\arraybackslash}m{0.25in}}
\newcolumntype{U}{>{\centering\arraybackslash}m{0.4in}}
\newcolumntype{Y}{>{\centering\arraybackslash}m{0.5in}}
\newcolumntype{Z}{>{\centering\arraybackslash}m{0.5in}}
\usepackage{hhline}

\def\BibTeX{{\rm B\kern-.05em{\sc i\kern-.025em b}\kern-.08em
    T\kern-.1667em\lower.7ex\hbox{E}\kern-.125emX}}
\begin{document}
\title{Transactive Control of Electric Railways Using Dynamic Market Mechanisms}
\author{David D'Achiardi, Anuradha M. Annaswamy, Sudip K. Mazumder, and Eduardo Pilo
\thanks{This work has been submitted to the IEEE for possible publication. Copyright may be transferred without notice, after which this version may no longer be accessible.} 
\thanks{This work was supported NSF’s Cyber-Physical Systems program under award 1644877 and 1644874. However, any opinions, findings, conclusions, or recommendations are those of the authors and do not necessarily reflect the views of the NSF.}
\thanks{D. D'Achiardi and A.M. Annaswamy are with the Department of Mechanical Engineering, Massachusetts Institute of Technology, Cambridge, MA, 02139 USA $\left[davidhdp, aanna\right]@mit.edu$.}
\thanks{E. Pilo is with Polytechnic School, Universidad Francisco de Vitoria, 28223 Pozuelo de Alarcón Madrid, Spain $eduardo.pilo@ufv.es$.}%
\thanks{S. K. Mazumder is with the Department of Electrical and Computer Engineering, University of Illinois at Chicago, Chicago, IL, 60607 USA $mazumder@uic.edu$.}}

\maketitle

\begin{abstract}
Electricity demand of electric railways is a relatively unexplored source of flexibility in demand response applications in power systems. In this paper, we propose a transactive control based optimization framework for coordinating the power grid network and the train network. This is accomplished by coordinating dispatchable distributed energy resources and demand profiles of trains using a two-step optimization. A railway based dynamic market mechanism (rDMM) is proposed for the dispatch of distributed energy resources (DER) in the power network along the electric railway using an iterative negotiation process, and generates profiles of electricity prices, and constitutes the first step. The train dispatch attempts minimize the operational costs of trains that ply along the railway, while subject to constraints on their acceleration profiles, route schedules, and the train dynamics, and generates demand profiles of trains and constitutes the second step. The rDMM seeks to optimize the operational costs of the underlying DERs while ensuring power balance. Together,
they form  an overall framework that yields the desired
transactions between the railway and power grid infrastructures.
This overall optimization approach is validated using simulation studies
of the Southbound Amtrak service along the Northeast
Corridor (NEC) in the United States, which shows a 25\%
reduction in energy costs when compared to standard trip
optimization based on minimum work, and a 75\% reduction
in energy costs when compared to the train cost calculated
using a field dataset.

\end{abstract}


\section{Introduction}
\label{sec:introduction}
\IEEEPARstart{M}{odern} electric trains can both demand power from their traction system for locomotion and inject power back into the electricity network through regenerative braking, virtually enabling them to store electricity in the form of kinetic energy  \cite{yu2007measurement}. The power profile of a train along a route is in many cases determined by the conductor based on training and experience, attempting to meet a given schedule with little regard to the varying electricity price along the route. In this paper, we propose an alternate operation methodology that consists of coordination of train schedules and the dispatch of  rail-side distributed energy resources (DERs) and leads to a determination of prices and schedules of power consumption for the trains and power generation for the DERs. Together, this coordinated operation is shown to minimize the overall electricity cost incurred by the trains. As this coordination occurs through a transactive framework between the train dispatch and the dispatch of railway agents such as DERs, this leads to a transactive control of the two interconnected systems of train network and the power grid network, similar to the transactive controller in  \cite{melton2015pacific, schweppe1980homeostatic}. 

A major driver that allows the proposed transactive coordination framework between the train-network and the power-network is the transformation of the latter in recent years. This has been due to the explosive growth of renewable energy sources \cite{iea2020global} and demand response \cite{iea2019tracking}, collectively denoted as DERs. The increasing footprint of these resources enables them to be dispatchable, and to enter into a transactional framework where incentive information (pricing) and commitment decisions (quantity) can be iteratively exchanged and arrived at an optimal solution. This iterative transactional framework is denoted as Dynamic Market Mechanisms (DMM), and has been explored in \cite{bejestani2014hierarchical,knudsen2015dynamic,shiltz2016practical,shiltz2016integrated}. The origins of transactive control are very much rooted within the energy application realm, as the ideas of using an incentive signal to alter the behavior of demand-side customers in power systems can be traced back to Schweppe's 1980 paper on homeostatic utility control \cite{schweppe1980homeostatic}. A large-scale demonstration of this concept can be found in \cite{melton2015pacific} where electric systems with high renewable adoption were considered and shown to meet demand
reduction objectives. The DMM-related results in \cite{bejestani2014hierarchical} show that DERs can engage in market-transactions at the tertiary control level and ensure grid objectives through a hierarchical framework. Exactly how these market mechanisms can be integrated into a real-time market and a regulation market together with secondary control based AGC was explored in \cite{shiltz2016integrated}. Focus was placed on demand response compatible assets in \cite{knudsen2015dynamic} . The implications of DMM in the context of a combined heat and power microgrid was explored in \cite{nudell2017dynamic}. In this paper, we will enable the coordination of various DERs in a power network that are located along the railway network through the use of a DMM, and denote it as rDMM.

The main challenge in the design of the proposed transactive framework is to coordinate the objectives of the two networks. The points of intersection between these two networks is the need to optimize operational costs and the need to ensure physics-based constraints such as power balance, capacity and operational limits, and train kinematics. A combined optimization problem subject to all underlying constraints can be posed and used to determine the train schedules and prices, but can prove to be quite intractable due to the complex nature of the constraints, space-dependent and time-dependent constraints with various intractable coupling mechanisms. We therefore adopt a two-step approach where the first consists of railway dispatch of schedules and electricity prices for a given train-demand profile, and the second consists of train dispatch which determines the train schedules for a given electricity-price profile. The railway dispatch determines, along each section of the track, the electrical output of each generator, the output of all storage assets, and the output of all cogeneration assets for a given set of profiles of power demand from trains and renewable generation. The train dispatch solves the trajectory optimization problem, i.e. the velocity profiles of the trains, through an energy cost minimization subject to acceleration limits and kinematic constraints. Our thesis in this paper is that such a two-step optimization can enable an effective coordination between the power grid network and railway network. In particular, we will show that the two-step optimization will lead to significant reduction in energy costs through simulation studies. While the methods utilized for solving this two-step optimization are fairly straightforward, the main contribution of the paper lies in the novelty of the proposed approach for trip optimization in electric railway networks. To our knowledge, such a transactive approach, which can be viewed as Demand Response using the flexibility in power consumption of trains, has not been suggested thus far in the literature except for \cite{dachiardi2019transactive} and  \cite{sarma2019costReduction}, where we presented preliminary results using this approach.

Related work that has addressed trip optimization in rail networks can be found in \cite{wang2011survey}, \cite{wang2016optimal}, \cite{khmelnitsky2000optimal}, \cite{franke2000algorithm}, and \cite{eldredge2011trip}. Reference \cite{wang2011survey} provides a summary of the trajectory planning problem in railway systems. Optimal trajectory planning for electric railways can be found in chapters 3 and 4 of \cite{wang2016optimal}, where pseudospectral methods are used to determine optimal railway operation based on models of train dynamics. This work builds upon the work minimization literature developed by \cite{khmelnitsky2000optimal} and \cite{franke2000algorithm}. Finally, \cite{eldredge2011trip} develops a control system to reduce fuel use in freight locomotives. In all of these lines of research, the overall objective is to minimize energy use, or work done by the train, rather than the cost of the electricity to the infrastructure manager, an important component of our proposed scheme.

The remainder of this paper is organized as follows. Section \ref{sec:ProblemFormulation} describes the problem faced by the railway and train operators in scheduling DERs and trains along the railway system. In Section \ref{sec:AgentDispatch} we break out the railway dispatch, i.e. dispatch of the generators and other assets along the railway track based on the estimated renewable generation, traction electric demand and electric and thermal loads for each section of the electric railway. Section \ref{sec:TrainDispatch} establishes the dynamic model of an electric train and formulates the energy cost minimization problem that needs to be solved by each train traveling along the electric railway. Section \ref{sec:TransArch} describes the integrated transactive control methodology that iterates between the Railway Dispatch and the Train Dispatch solutions to determine the price signals and corresponding dispatch profiles of the agents and trains that minimize the operational costs of the entire system. Section \ref{sec:Simulations} presents a case study of the Amtrak service along the Northeast Corridor in the United States, composed of multiple sections and DER topologies, and validate the proposed transactive controller. Realistic accommodation of data has been carried out in this case study including incorporation of the actual electricity prices from the wholesale market, actual load profiles, realistic train data, and renewable energy profiles available in public database. Using this case study, we compare our approach with both the current train profile using field data and an optimization framework based on minimization of work. Concluding remarks and future research extensions are discussed in Section \ref{sec:Conclusion}. 

\section{Problem Formulation}
\label{sec:ProblemFormulation}
Electric railway systems can be owned and operated by multiple parties. In some systems, the track and electric system that powers the trains are owned by an entity that charges a fee for the utilization of their facilities. This entity is typically responsible for the maintenance of the track, procuring the electric power to feed the trains, dispatching rail-side DERs and controlling traffic along the system. These entities are commonly known as Railway Operators. We will denote Train Operators as those who are in charge of the use of the track based on their projected train schedules, maintain and dispatch the trains and pay usage fees to the Railway Operator. The problem that we consider in this paper is a combined optimization of both the railway dispatch and the train dispatch, with Railway Operator and Train Operator acting as the interface between the grid network and the train network (see Figure \ref{fig:rDMMdiagram} for an overall schematic).

In Section \ref{subsec:ProblemFormulation-EnergyProcurement} we will discuss how Railway Operators procure electricity and thermal energy from wholesale energy markets, distribution operators (i.e. utility companies) and DERs. In Section \ref{subsec:ProblemFormulation-SocialWelfareMax} we formulate the underling optimization problem where the cost is Social-Welfare like, and depends on the costs incurred by the DERs, the electrical and thermal loads of the railway system and the electric trains. The resulting solution will then provide optimal profiles for various generation assets as well as the train consumption/generation profiles, i.e. the railway dispatch and the train dispatch. Deriving such a solution however proves exceedingly difficult to solve due to the presence of several dynamic nonlinear constraints and the fact that the timescale required to solve the trajectory optimization problem needs to approach real-time, especially for the train dispatch. This motivates the two part solution proposed in Sections \ref{sec:AgentDispatch}-\ref{sec:TransArch}.

\subsection{Traction System Preliminaries}
\label{subsec:ProblemFormulation-EnergyProcurement} 
A wide range of traction system architectures and technologies have been developed for electrified railway systems, across low frequency (DC, 16.6Hz, 20 Hz and 25Hz) as well as industrial frequency (50Hz and 60Hz) networks which are powered by overhead and third rail traction distribution systems. Additionally, electric trains have been both powered by rail-side generators as well as grid-tied systems. In some scenarios, electrical connections to a large-scale electricity distribution or transmission system are complemented with distribution lines that travel along the rail, and can be designed such that they improve the reliability of the traction system by providing redundant power supplies to the traction substations feeding the traction system. These various architectures have a direct impact on the flexibility of the demand of the train and therefore a possible energy cost reduction.

\begin{figure}
	\centering
	\includegraphics[width=3.5in]{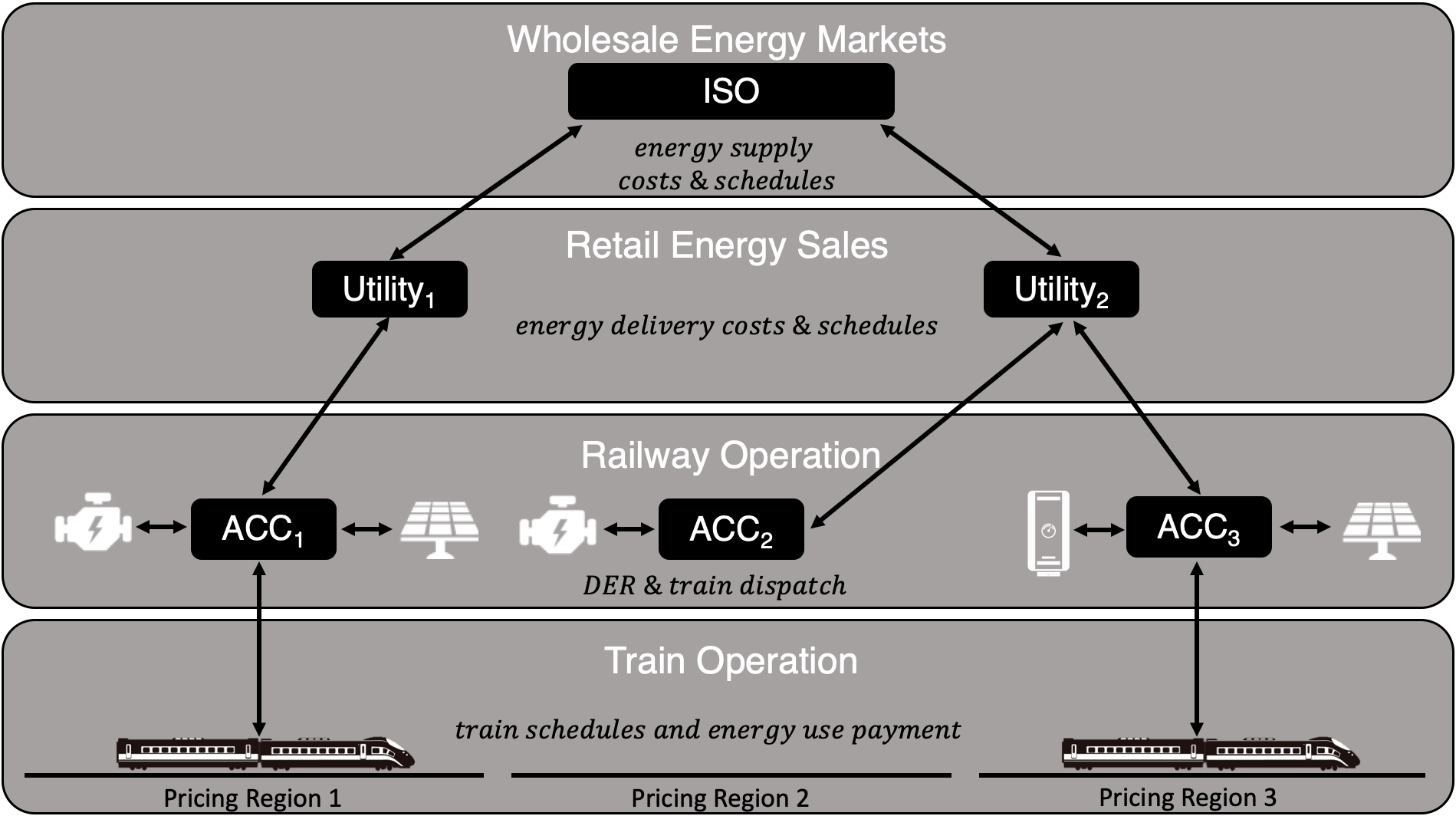}
	\caption{Summary of the existing activities (power, cost and information flows) between the four entities (wholesale energy markets, retail energy sales, railway dispatch and train dispatch) to power within the electric railway network.}
	\label{fig:SystemOrganization}
\end{figure}

A typical architecture of energy procurement and dispatch that occurs along the electric railway is provided in Fig. \ref{fig:SystemOrganization} and will be adopted for the discussions in this paper. This architecture is typically adopted in deregulated electricity markets such as Amtrak in the Northeastern United States, and will be used in the case study presented in Section \ref{sec:Simulations}. That is, we assume that Railway Operators face delivery charges as distribution-level customers of electric utilities which own and operate the distribution system. However, due to their energy requirements and the regulatory setting in which they procure energy, they could access wholesale markets for electrical supply. Additionally, we assume that third-parties could invest in rail-side DERs, seeking to provide energy services to the Railway Operator. Competitive retail sales of electricity from power marketers \cite{manzagol2018powerMarketers} and on-site Power Purchase Agreements (PPAs) offered by Energy Services COmpanies (ESCOs) \cite{stuart2016us} are both established and growing energy procurement mechanisms for commercial companies including electric railways.

\subsection{Overall Constrained Optimization Problem}
\label{subsec:ProblemFormulation-SocialWelfareMax} 
With the architecture chosen as in Fig. \ref{fig:SystemOrganization}, we now introduce a problem formulation for energy procurement and dispatch for optimal grid-railway interaction. A first component of this problem formulation is the Area Control Centers(ACCs) (see Fig. \ref{fig:SystemOrganization}) managed by a Railway Operator. An ACC is charged with serving the electric railway traction system, limited to a portion of contiguous electric railway with the property that marginal injections or demands for power have the same cost to the operator for all time $t$. Along this portion of the track, all rail-side DERs and trains that interface with the traction system are enabled to provide price and quantity information regarding their dispatch, and are compensated and charged based on their actions by the ACC. The Railway Operator manages the ACC with the objective of reducing overall operational costs while maximizing the value of the DERs along the track and meeting all thermal and electric loads in the system. That is, the overall objective of the railway dispatch is to schedule energy resources, which includes trains, so as to minimize the cost of energy resources and trains along the railway system. Such an optimization has to be carried out subject to the various constraints of the grid and railway networks. 

More formally, the underlying problem is the dispatch of electric railway power systems across $N$ ACCs for a time horizon $\tau = \left[t_0,t_f\right]$ in an optimal manner. This includes the electric traction loads of $L_n$ trains, with power profiles denoted as $P_n^l$ for the electric profile of train $l$ in $ACC_n$ over the time interval $\left[ t_{n,0}^l, t_{n,f}^l\right]$. The Railway Operator is charged for its electric loads at a rate $\lambda_n\left(t\right)$.

On the generation side, we consider $D_n$ dispatchable generators (including DERs and electric utility imports) at each $ACC_n$ with electric and thermal power profiles $P_n^{d_e}$ and $P_n^{d_{th}}$. For compactness, we will use $P_n^{d}$, a tuple of the electric and thermal generation for each generator $d$. These agents incur a cost $C_n^d\left(P_n^d\right)$, which is private to their owner and operator. 

All other electric and thermal power demand in the electric railway system, such as the thermal conditioning and lighting loads at passenger stations, are considered price-inelastic and are included within the power balance constraints of the problem but are not included in the objective function due to their fixed nature.

Given that the utility of the trains and generators are fixed, the social welfare maximization problem, which is commonly used for economic dispatch problems in power grids, reduces to a cost minimization problem with two terms in the objective function, the generator cost of generating the electricity and the energy cost of operating the trains.

With the above definitions, the overall grid-railway optimization can be posed thus:

\begingroup
\allowdisplaybreaks
\begin{align}
& \underset{P^D_n(t), P^T_n(t)}{\text{min}} &&\sum^N_{n=1}\Biggl( \sum^{D_n}_{d=1}C_n^d(P_n^d) \label{eq:ProblemObjective}\\
&&&\qquad +{\sum^{L_n}_{l=1}{ \int_{\tau=t_{n,0}^l}^{\tau=t_{n,f}^l} P^l_n \left( \tau \right)\lambda_{n} \left( \tau \right) d\tau}} \Biggr) \nonumber \\
& \text{s.t.} &&\text{Nodal Electric Power Balance}\label{eq:ProblemConstraintElectricPowerBalance}\\ 
&&&\text{Nodal Thermal Power Balance}\label{eq:ProblemConstraintThermalPowerBalance}\\ 
&&&\text{DER - Agent Capacity Limits}\label{eq:ProblemConstraintAgentCapacityLimits}\\ 
&&& \text{Train Dynamics}\label{eq:ProblemConstraintDynamics}\\
&&&\text{Train Electric Motor Power Limits} \label{eqProblemConstraintPower}\\
&&&\text{Train Traction Force Limits} \label{eq:ProblemConstraintTractionForce}\\
&&&\text{Train Acceleration Limits} \label{eq:ProblemConstraintAcceleration}\\
&&&\text{Train Velocity Limits} \label{eq:ProblemConstraintVelocity}\\
&&&\text{Train Schedules} \label{eq:ProblemConstraintSchedule}.
\end{align}
\endgroup

\noindent We note that in the constrained optimization problem in Equations (\ref{eq:ProblemObjective})-(\ref{eq:ProblemConstraintSchedule}), the decision variables are composed of the power profiles of the dispatchable agents and the trains at each of the ACCs as well as the price of electricity paid by the electric trains.

It is clear that the objective function in Eq. (\ref{eq:ProblemObjective}) is composed of the sum of the cost of the dispatchable generators (which we will denote as agents) and the energy expenses of the electric trains. These two costs are the fundamental building blocks of the two part railway dynamic market mechanism ($rDMM$) developed in Sections \ref{sec:AgentDispatch}-\ref{sec:TransArch}. When broken down into the individual ACCs, the first term captures the railway dispatch problem with fixed loads, which in turn is subject to power balance and agent capacity limit constraints (\ref{eq:ProblemConstraintElectricPowerBalance})-(\ref{eq:ProblemConstraintAgentCapacityLimits}). The second term corresponds to the train dispatch problem, and is in turn subject to train dynamics and scheduling constraints (\ref{eq:ProblemConstraintDynamics})-(\ref{eq:ProblemConstraintSchedule}).

The optimization problem as in (\ref{eq:ProblemObjective})-(\ref{eq:ProblemConstraintSchedule}) is difficult to solve, highly intractable, and poses several challenges. First, its solution requires that all of the traction system agents share their private cost information $C\left(P_n^d\right)$ to appropriately capture and minimize the overall system cost. This feature would likely inhibit private investment in rail-side DERs. The timescale of electric train dispatch may not individually align with that of the $ACC_n$ dispatch. Trajectory optimization of electric railway power profiles occurs in the seconds timescale whereas energy asset dispatch is unlikely to be used at a timescale faster than 5 minutes. Additionally, the trains are entering and exiting each of the $ACC_n$ at different times which are dependent on the pricing signal. All of these cause an intricate coupling between spatial and temporal constraints that are nonlinear and convex. In an effort to simplify the process and make the problem more tractable, we propose a two step approach that iterates between the electric railway and the train dispatch problems in Sections \ref{sec:AgentDispatch} - \ref{sec:TransArch}.

\section{Electric Railway Dispatch}
\label{sec:AgentDispatch}
This section describes the minimization of energy asset costs in an electric railway where the Railway Operator must guarantee that all electric and thermal constraints must be met. We will explicitly accommodate the constraints and timescales of each energy agent along the railway. Both thermal and electrical energy assets powering the electric railway are included in this optimization. Section \ref{subsec:AgentDispatch-AgentTypesTimescales} describes the different agents that operate in the system and introduces the timescales in which they interact. Section \ref{subsec:AgentDispatch-AgentCostsOperation} discusses the model used for agent costs and operations within the Railway Dispatch problem. Lastly, Section \ref{subsec:AgentDispatch-Algorithm} uses the agent costs and operations to state and solve the Railway Dispatch problem, following a similar procedure as the one developed in \cite{nudell2017dynamic}.

\subsection{Agent Types and Timescales}
\label{subsec:AgentDispatch-AgentTypesTimescales}
Within each of the railway segments $n$ the Railway Operator must meet the electrical and thermal demand or load during the next $M$ future dispatch intervals which are indexed as $K=\{1,...,M\}$. Moreover, in managing the electrical system of the railway, the Railway Operator is tasked with dispatching and compensating the electrical agents or assets within $ACC_n$ in a way that minimizes the total cost of the operation. The electric railway dispatch problem described in this section determines optimal dispatch for each future time interval; however, the pricing and dispatch is only binding and executed for $K = 1$ after the Railway Dispatch problem has been solved.

\begin{figure}
	\centering
	\includegraphics[width=3.5in]{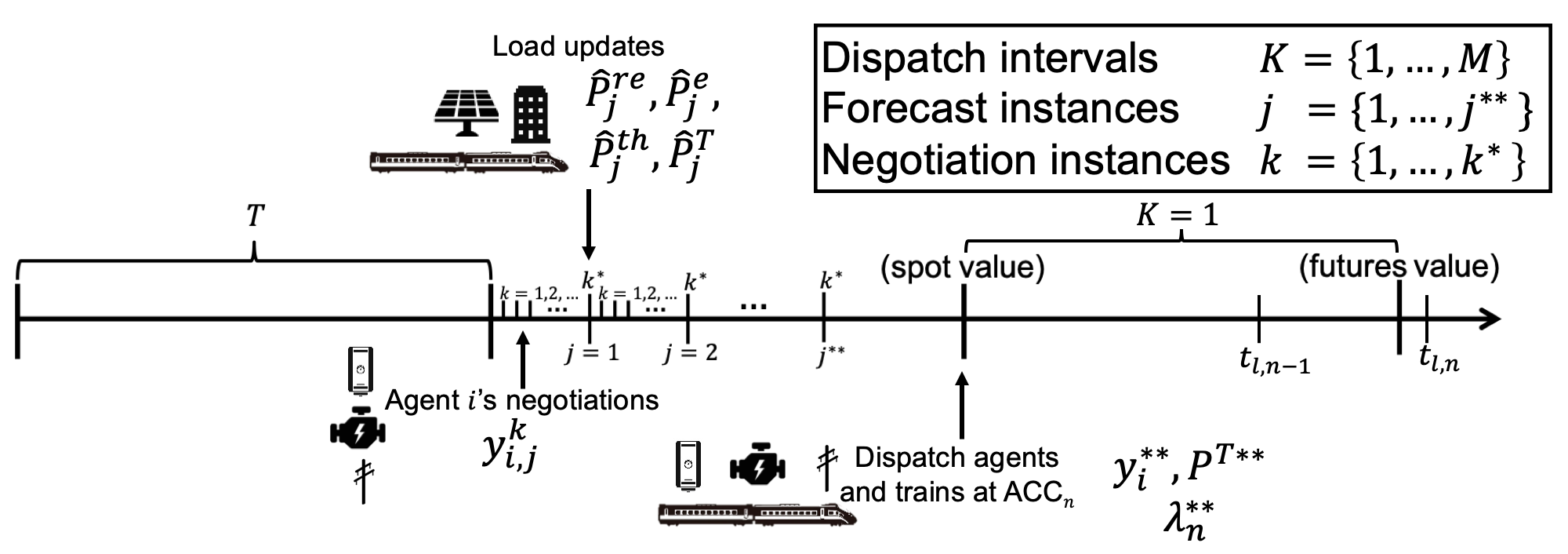}
	\caption{Time scales within the Railway Dispatch problem. Dispatch intervals $K = \{1,...,M\}$ constitute the rolling dispatch horizon considered by the Railway Operator of $ACC_n$. The passive agents $\mathcal{F}_n$ and electric trains $\mathcal{T}_n$ provide a thermal and electric load forecast for each dispatch interval during forecast instances $j = \{1,...,j^{**}\}$. Active agents $\mathcal{A}_n$ negotiate price and quantity during each negotiation instance $k=\{1,...,k^*\}$ based on the forecast updates and their private cost information. It is assumed that train $l$ traverses $ACC_n$ over the period $[t_{l,n-1},t_{l,n}]$ which overlaps with the forecast intervals $K=1$ and $K=2$.}
	\label{fig:rDMMtimeline}
\end{figure}

We define five types of dispatchable agents within the railway power system at each node $n$ that are classified into the following sets: heating assets (e.g. boilers), $\mathcal{H}_n$; electric generation assets (e.g. fuel cell, microturbines), $\mathcal{E}_n$; cogeneration assets (e.g. combined heat and power units), $\mathcal{C}_n$; storage assets (e.g. batteries), $\mathcal{S}_n$; low voltage side network connections (i.e. points of common coupling), $\mathcal{N}_n$. The set of all dispatchable agents at node $n$ is denoted as $\mathcal{A}_n \triangleq \mathcal{H}_n \cup \mathcal{E}_n \cup \mathcal{C}_n \cup \mathcal{N}_n$. It is assumed that within each dispatch interval $K$ there are two faster timescales, where the first corresponds to instances $j\in \{1,...,j^{**}\}$, at each of which forecast updates of all non-dispatchable loads and generation are received. Between each $[j, j+1]$, we introduce a faster timescale $k\in \{1,...,k^*\}$ where at each instance $k$, all dispatchable agents negotiate electric and thermal generation schedules and prices (see Fig. \ref{fig:rDMMtimeline}). It is assumed that these timescales with $j^{**}$ and $k^*$ are such that they are nested, and that the time-intervals permit a useful forecast data and sufficient negotiations, respectively. 

In addition to the above agents, we also consider three types of non-dispatchable agents at each node $n$, classified into the following sets: renewable generators, ${re}_n$; electrical loads (e.g. lighting at the passenger station), ${e}_n$; and thermal loads (e.g. heating of the passenger stations), ${th}_n$. The set of these non-dispatchable agents at node $n$ is denoted as $\mathcal{F}_n \triangleq {re}_n \cup {e}_n \cup {th}_n$, all of whom inject and demand electric and thermal energy from the same network as the dispatchable agents but are not dispatched by the Railway Operator. Instead, the operators in charge of each of these passive resources (i.e. the renewable asset operator for ${re}_n$ and the passenger station operator for $e_n$ and ${th}_n$) communicate the best load estimate over $ACC_n$'s future time intervals $K$. This forecast update occurs every $j$ when new updates arrive for the renewable assets, or any of the loads. The periods associated with the time scales $k$ and $j$ need to be such that it should accommodate new information related to the forecast and at the same time, the convergence rates of the negotiations.

Lastly, we consider electric trains $l = \{1,...,L\}$ injecting and demanding electric power from $ACC_n$. We denote the set of all trains at node $n$ as $\mathcal{T}_n$. For the purpose of the Railway Dispatch problem, the trains are no different from the set of passive agents $\mathcal{F}_n$ in that they provide an update to their electric power profile for the dispatch intervals $K=\{1,...,M\}$ at each forecast instance $j$. However, as shown below, once we define the Train Dispatch problem faced by the train operators for each train in Section \ref{sec:TrainDispatch} and the composition of the transactive controller in Section \ref{sec:TransArch}, the train update will transform from a simple forecast update to a price-dependent minimum cost dispatch update.

\subsection{Agent Costs and Operation}
\label{subsec:AgentDispatch-AgentCostsOperation}
The overall goal of Railway Dispatch is to arrive at pricing information that can yield optimal setpoints for generators and utility power import. Over the time horizon of $M$ dispatch intervals, the electric and thermal power profiles for each of the agents considered (dispatchable, non-dispatchable, and trains) is assumed to take positive (generation) and negative (load) values.

In order to capture the electric and thermal components of generation for dispatchable agents in $\mathcal{A}_n$ at each time period $M$, the output of the generators is denoted by the decision variable
\begin{align}
y_i \in \R^M \qquad \forall \; i \; \in \; \mathcal{A}_n
\label{eq:AgentDecisionVariable}
\end{align}
and the $K$'th element of $y_i$ as $y_{i,K}$.This decision variable maps to an electric and a thermal output as given by:
\begin{align}
g_i^e(y_i) = d_i^ey_i \qquad \forall \; i \; \in \; \mathcal{A}_n,
\label{eq:AgentElectricConversion}
\end{align}
and 
\begin{align}
g_i^{th}(y_i) = d_i^{th}y_i \qquad \forall \;  i \;  \in \; \mathcal{A}_n.
\label{eq:AgentThermlConversion}
\end{align}
\noindent In other words, $y_i$ is a dispatch setpoint associated with a particular electric and thermal output. For electric-only generator $i$, the thermal conversion coefficient vector, $d_i^{th} \in \R^M$, is the zero vector and the electric conversion coefficient vector, $d_i^{e} \in \R^M$ takes positive values. We make the following assumptions regarding their constraints and costs.

\begin{assumption} Electric and thermal power capacity of agent $i$ are bound by $\{\underline{P_i^e},\overline{P_i^e}\}$ and $\{\underline{P_i^{th}},\overline{P_i^{th}}\}$ respectively.
\end{assumption}

\begin{assumption} Capacity constraints are not binding and losses are negligible for all electrical and thermal equipment in the system other than the dispatchable agents.
\end{assumption}

\begin{assumption}
The cost function of each dispatchable agent $i \in A_n$ is a convex quadratic function. 

For a single dispatch interval $K = \{1,...,M\}$, this quadratic cost is denoted as:
\begin{align}
J_{i,K}(y_{i,K}) = a_{i,K} + b_{i,K}y_{i,K} + \frac{1}{2}c_{i,K}y_{i,K}^2
\label{eq:AgentCostFunction}
\end{align}
\end{assumption}
and over the multi-periods as 
\begin{align}
J_i(y_i) = \sum_{K=1}^{M}J_{i,K} \left(y_{i,K} \right) \qquad \forall i \; \in \; \mathcal{A}_n.
\label{eq:AgentCostCurve}
\end{align}
For the agents representing low voltage side network connections $\mathcal{N}$, the cost function can be updated as a function of the equilibrium price in an external market, such as a wholesale energy market. Labeling the external market price for the low voltage side network connections as $\pi^\mathcal{N}_{n,j}$, the cost function parameters $a_{i,K}$, $b_{i,K}$, and $c_{i,K}$ in (\ref{eq:AgentCostFunction}) are determined for $i\in\mathcal{N}$ at each forecast instance $j$. Given that these market prices commonly represent marginal prices, the cost function parameters may be simply chosen as: $a_{i,K}=0$, $b_{i,K}=\pi^i_{n,j}$, and $c_{i,K}=0$.
 
Each of the passive agents in $\mathcal{F}_n$ determine their power output for the dispatch intervals $K = \{1,...,M\}$ at each forecast instance $j$ (see Fig. \ref{fig:rDMMtimeline}), which are denoted by $\hat{P}^{re}_{j,K}$ for renewable generators, $\hat{P}^e_{j,K}$ for electric loads and $\hat{P}^{th}_{j,K}$ for thermal loads. We drop the subscript $K$ to denote the power output vector over the dispatch intervals in $\R^M$ as $\hat{P}^{re}_{j}$, $\hat{P}^e_{j}$, and $\hat{P}^{th}_{j}$. In order to determine the train demand, suppose that train $l$ traverses $ACC_n$ over the period $[t_{l,n-1},t_{l,n}]$ over the dispatch interval $K$ and $P_l(t,n)$ is the corresponding forecasted demand at instance $j$, we denote this demand as $P_{l,j,K}^*$. This in turn can be summed over all $L$ trains to yield 
\begin{align}
\hat{P}_{j,K}^T = \sum_{l=1}^{l=L}{P_{l,j,K}^*}.
\label{eq:AgentTractivePower}
\end{align} 
For ease of exposition, we drop the subscript $K$ in (\ref{eq:AgentTractivePower}) and simply denote the total train demand at forecast instance $j$ as $\hat{P}_{j}^T$.

\subsection{Railway Dispatch Algorithm}
\label{subsec:AgentDispatch-Algorithm}
With the overall costs and constraints related to all agents specified as above, we state the Railway Dispatch problem at $ACC_n$ at a fixed forecast instance $j = \{1,...,j^{**}\}$ over the dispatch intervals $K = \{1,...,M\}$ in $\R^M$ as:

\begin{align}
&\underset{y_{i,j} \; \forall \; i \; \in \; \mathcal{A}_n}{\text{min}}&&\sum_{i\in\mathcal{A}_n}J_i\left(y_{i,j}\right) \label{eq:AgentDispatchObjectiveCostMinimization}\\
&\text{s.t.} && c_e = \hat{P}^{re}_{j}+\hat{P}^T_{j}+\hat{P}^e_{j}+\nonumber \\
&&& \qquad \sum_{i \in \mathcal{A}_n}g_i^e(y_{i,j})=0 \label{eq:AgentDispatchConstraintElectricBalance}\\
&&&c_{th}=\hat{P}_{j}^{th}+\sum_{i \in \mathcal{A}_n}g_i^{th}(y_{i,j})=0 \label{eq:AgentDispatchConstraintThermalBalance}\\
&&& m_i^+(y_{i,j})=y_{i,j}- \overline{y_{i,j}}\leq0\label{eq:AgentDispatchConstraintCapacityUpperBound}\\
&&& m_i^-(y_{i,j})=\underline{y_{i,j}}-y_{i,j}\leq0. \label{eq:AgentDispatchConstraintCapacityLowerBound}
\end{align}
 
The output values for each dispatchable agent $i \in \mathcal{A}_n$ over the dispatch interval $K = \{1,...,M\}$ at dispatch instance $j$ are denoted as $y_{i,j} \in \R^M$ and constitute the decision variables of the problem. These decision variables are bound at each dispatch interval $K$ by the sum of the electric loads per (\ref{eq:AgentDispatchConstraintElectricBalance}), the thermal loads per (\ref{eq:AgentDispatchConstraintThermalBalance}), the maximum capacity constraint per (\ref{eq:AgentDispatchConstraintCapacityUpperBound}) where $\overline{y_{i,j}} = min\{\overline{P_i^e}/d_i^{e}, \overline{P_i^{th}}/d_i^{th}\}$ and the minimum capacity constraint per (\ref{eq:AgentDispatchConstraintCapacityLowerBound}) where $\underline{y_{i,j}} = max\{\underline{P_i^e}/d_i^{e}, \underline{P_i^{th}}/d_i^{th}\}$.

At each forecast instance $j = \{1,...,j^{**}\}$, the power profile for each renewable generator, $\hat{P}^{re}_{j} \in \R^M$, electric load, $\hat{P}^e_{j}  \in \R^M$, thermal load, $\hat{P}^{th}_{j}  \in \R^M$, and the total traction load, $\hat{P}_{j}^T  \in \R^M$, are updated for the dispatch interval $K = \{1,...,M\}$. The new forecasts are used to update constraints (\ref{eq:AgentDispatchConstraintElectricBalance}) and (\ref{eq:AgentDispatchConstraintThermalBalance}), with the resulting optimization problem in (\ref{eq:AgentDispatchObjectiveCostMinimization})-(\ref{eq:AgentDispatchConstraintCapacityLowerBound}) solved again.

With each forecast update $j=j+1$, the decision variables of the problem, denoted as $y_{i,j}^*$ optimize the cost in (\ref{eq:AgentDispatchObjectiveCostMinimization}). For the dispatch intervals $K = \{1,...,M\}$ these decision variables can in turn be mapped to the electric and thermal output of the agents as $[g_i^{e}(y_{i,j}^*),g_i^{th}y_{i,j}^*)]\in \R^M$. 

The underlying optimization problem that the Railway Operator has to solve is therefore given by the solution of (\ref{eq:AgentDispatchObjectiveCostMinimization})-(\ref{eq:AgentDispatchConstraintCapacityLowerBound}) for a given set of forecasted profiles. We propose that each $ACC_n$ solves this through an iterative negotiation process amongst the dispatchable agents within this $ACC$. This is proposed to be carried out by the faster timescales, $k=1,2,... k^*$ for each $j$. That is, at each forecast instance $j$ the negotiation process starts at $k=0$, where $\hat{P}^{re}_{j}$, $\hat{P}^e_{j}$, $\hat{P}^{th}_{j}$, and $\hat{P}_{j}^T$ are fixed, allowing the Railway Operator at $ACC_n$ to establish the electric and thermal loads for the dispatch horizon. And at $k = 0$, it is assumed that the Railway Operator for $ACC_n$ broadcasts an initial price duple $\lambda_{n,j,k=0} = \left[\lambda^e_{n,j,0},\lambda^{th}_{n,j,0}\right] \in \R^M$ consisting of electric and thermal prices.

With these initial conditions, using a Lagrangian and a gradient-based update, the decision variables $y_{i,j}^k \in \R^M$ and the prices $\lambda^e_{n,j,k}\in \R^M$ and $\lambda^{th}_{n,j,k}\in \R^M$ are updated at each negotiation instance $k$ as:

\begin{align}
y_{i,j}^{k+1} = y_{i,j}^k - \beta_{y_{i}}\Bigl(\nabla_{y_{i,j}^k}J_i(y_{i,j}^k)+\left[\nabla_{y_{i,j}^k}h^e\right]^T\lambda^e_{n,j,k} \nonumber\\
-\left[ \nabla_{y_{i,j}^k} h^{th} \right]^T \lambda^{th}_{n,j,0} \mp \mu_{i,j}^{\pm k} \Bigr) 
\label{eq:AgentOfferUpdate}
\end{align} 
\begin{align}
\lambda^e_{n,j,k+1} = \lambda^e_{n,j,k} + \beta_{\lambda^e}(\hat{P}^{re}_{j}+\hat{P}_{j}^T+\hat{P}^e_{j}\nonumber \\
+\sum_{i \in \mathcal{A}_n}g_i^e(y_{i,j}^k)) \label{eq:AgentElectricPriceUpdate}\\
\lambda^{th}_{n,j,k+1} = \lambda^{th}_{n,j,k} + \beta_{\lambda^{th}}(\hat{P}^{th}_{j}+\sum_{i \in \mathcal{A}_n}g_i^{th}(y_{i,j}^k))
\label{eq:AgentThermalPriceUpdate}
\end{align} 
where $\beta_{y_{i}}, \beta_{\lambda^e}, \beta_{\lambda^{th}} \in \R \; \forall \; i \in \mathcal{A}_n$ are positive step-size parameters and $ \mu_{i,j}^{\pm} \in \R \; \forall \; i \in \mathcal{A}_n$  is the penalty for violating capacity constraints (\ref{eq:AgentDispatchConstraintCapacityUpperBound}) and (\ref{eq:AgentDispatchConstraintCapacityLowerBound}). The penalty function updates are given by:

\begin{align}
\mu_{i,j}^{\pm k+1} = max\{0\;,\;\mu_{i,j}^{\pm k}+\beta_{\mu_{i}}\mu_{i,j}^{\pm k}\}
\label{eq:AgentPenaltyFunctionUpdate}
\end{align}
\noindent where $\beta_{\mu_{i}}$ is a positive step-size parameter.

It is assumed that these iterations occur at each $k$ and converge as $k \rightarrow k^*$ for some $k^*$. As outlined in \cite{shiltz2016integrated}, under suitable convexity conditions, it can be shown that convergence to unique optimal values takes place.

Defining $\hat{P}^{F}_{j} = [\hat{P}^{re}_{j}+\hat{P}^e_{j},\hat{P}^{th}_{j}]$ as the estimate of the fixed assets at $j$, these estimates can be updated with $j$ as the new forecasts arrive. The forecast update enables an improved dispatch of the agents across the dispatch intervals $K=\{1,...,M\}$.

Note that in practice, exit conditions based on the agent output and price profile updates within the negotiation process can be established such that the forecast update is promptly initiated. These exist conditions follow the form: $\left|y_{i,j}^{k+1}-y_{i,j}^k\right|\leq\gamma_k^y \; \& \; \left|\lambda_{n,k+1,j}-\lambda_{n,k,j}\right|\leq\gamma_k^\lambda $. If these conditions are met, the equilibrium agent output and price profiles can be set as $y_{i,j}^{*} = y_{i,j}^{k+1} \forall i \in \mathcal{A}_n,$ and $\lambda_{n,j}^{*}=\lambda_{n,k+1,j}$ respectively.

Similarly, exit conditions can be established for the forecast update process, such that the agent output and price profile updates can be communicated for dispatch and settlement if so desired. These exit conditions follow the form: $\left|y_{i,j+1}^*-y_{i,j}^*\right|\leq\gamma_j^y \; \& \; \left|\lambda_{n,j+1}^*-\lambda_{n,j}^*\right|\leq\gamma_j^\lambda$. If these conditions are met, dispatch for the first interval $K=1$ can be established using agent output profiles $y_i^{**} = y_{i,j+1}^* \forall i \in \mathcal{A}_n$ and price profiles $\lambda_{n}^{**}=\lambda_{n,j+1}^*$. 

In summary, the Railway Dispatch algorithm for $K=1$ starts at $k=1, j=1$ with a forecast of the power profile for each renewable generator, $\hat{P}^{re}_{j} \in \R^M$, electric load, $\hat{P}^e_{j}  \in \R^M$, thermal load, $\hat{P}^{th}_{j}  \in \R^M$, and the total traction load, $\hat{P}_{j}^T  \in \R^M$, and returns the optimal agent output profiles $y_i^{**}$, the total traction demand profile $P^{T**}$ and the price profiles $\lambda_{n}^{**}$ that can be used for dispatch and settlement of dispatch interval $K=1$. The dispatch interval window then shifts over, with $K=2$ corresponding to the active dispatch interval and the process repeats. 

\section{Train Dispatch}
\label{sec:TrainDispatch}
In the previous section, we assumed that at each $ACC_n$, the train loads $\mathcal{T}_n$ were fixed, and the dispatch of the active agents $A_n$ was optimized. In this section, we address the fact that these train loads are indeed flexible, and pose a constrained optimization problem for determining the optimal profiles of generation and consumption of their power profiles. In Section \ref{subsec:TrainDispatch-DynamicModel}, we describe the physical model of the train. 
Next, we define the cost minimization problem for a track with multiple pricing regions, managed by $ACC_n$ in Section \ref{subsec:TrainDispatch-MinCostFormulation}. With this Train Dispatch accomplished, in Section \ref{sec:TransArch}, we describe how the Railway Dispatch optimization described in Section \ref{sec:AgentDispatch} can be stitched together with the Train Dispatch optimization to result in an overall transactive control framework for the combined grid-railway infrastructure. This overall framework is validated using numerical simulation of the Amtrak NEC in Section \ref{sec:Simulations}.

\subsection{Dynamic Model of Electric Trains}
\label{subsec:TrainDispatch-DynamicModel}
\begin{figure}[tp]
	\centering
	\smallskip
	\includegraphics[width=3.5in]{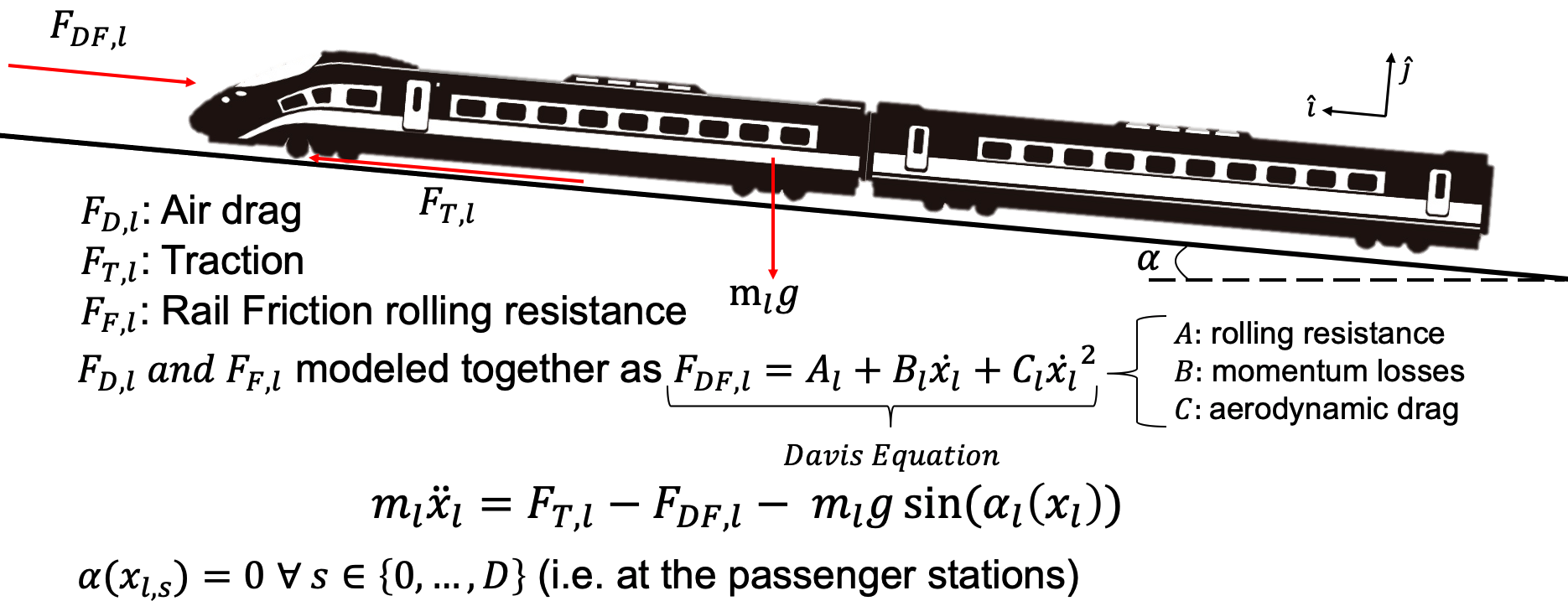}
	\caption{Free body diagram of electric train $l$. The resulting traction force $F_{T,l}$, friction and drag force $F_{DF,l}$ and the gravitational force component in the direction of motion of the train $m_lgsin\alpha_l(x_l)$ are identified. Newton's second law of motion is written for the $\ihat$ direction, along the direction of motion of the train.}
	\label{FBD}
\end{figure}

The power consumption of the electric train depends on the traction force, which in turn depends on the overall train dynamics. In this section, we derive the underlying dynamic model of the train and the corresponding power consumption profile. The position, velocity and acceleration of train $l$ in the direction of motion $\ihat$ are denoted by $x_l$, $\dot{x_l}$, and $\ddot{x_l}$ respectively. We proceed by defining the three forces with a component in the direction of motion of the train $\ihat$ as it travels at an angle $\alpha_l(x_l)$ from the horizon. The gravitational force on the train can be decomposed into the direction of motion $\ihat$ as $-m_lgsin(\alpha_l(x_l))$ and the direction normal to the ground $\jhat$ as $-m_lgcos(\alpha_l(x_l))$, where $m_l$ is the total mass of the train.

The electric motors converting electrical into mechanical power is assumed result in the traction force $F_{T,l}$ in the $\ihat$ direction. An opposing force $F_{DF,l}$ is also included in the model, which includes a drag force and a friction force \cite{davis1926tractive}. With these definitions, we can derive the equation of motion in the direction of motion $\ihat$ as 
\begin{align}
m_l\ddot{x}_{l} = F_{T,l} - F_{DF,l} - m_lgsin\alpha_l(x_l).
\label{eq:TrainDynamics}
\end{align}
The traction force $F_Tl$ is a function of both the electric power $P_l$ and $\dot x_l$, and is in general a mapping of the ratio $P_l/\dot x_l$. The drag-friction force can be represented as
\begin{align}
\sum{F_{DF,l}} = A_l + B_l\dot{x_l}+C_l\dot{x_l}^2.
\label{davisEquation}
\end{align}

\noindent known as Davis Equation \cite{davis1926tractive}. This industry standard approximation captures the rolling resistance effect at low speed through the linear term and the drag force through the quadratic term. More detailed drag-friction models for train cars can be found in \cite{hay1982railroad}.

\subsection{The Constrained Optimization Problem}
\label{subsec:TrainDispatch-MinCostFormulation}
In the analysis henceforth, we consider the cost minimization problem of train $l$ that departs location $x_0$ at time $t_0$ and arrives at a final destination $x_f$ at time $t_f$. The train stops at passenger stations denoted by $s \in \{0, ... , D\}$ between $t\in[t_{l,a}(s),t_{l,d}(s)]$, where $t_{l,a}(s)$ is the arrival time and $t_{l,d}(s)$ is the departure time from station $s$ and $x_{l,s}$ denotes the position of station $s$. And as indicated by (\ref{eq:TrainDynamics}), $x_l$ is determined by the train dynamics.

The train trajectory is assumed to traverse $n$ subsections, each dispatched by an Area Control Center, denoted as $ACC_n, \forall n \in \{1,...,N\}$. It should be noted that these sections may not coincide with the railway sections between passenger stations. Section $n$ of the track is bound by the position interval $[x_{n-1},x_n]$ and, as mentioned earlier, train $l$ is assumed to travel within this section during the time interval $[t_{l,n-1},t_{l,n}]$ for $ACC_n$. The operator of train $l$ must choose the electric power demand profile given by the set of functions $P_l(t,n)$, over this interval, $\;\forall\; n \in \{1,...,N\}$. The power demand profile of train $l$ is therefore composed of the profiles at each ACC:
\begin{align}\label{eq:TrainPowerProfile}
P_l(t)=\begin{cases} 
P_l(t,1) & t\in[t_{l,0},t_{l,1}]\\
&\vdots\\
P_l(t,N) & t\in[t_{l,N-1},t_{l,N}].
\end{cases}
\end{align}
We assume that the railway is level at the passenger stations (i.e. $\alpha(x_{l,s}) = 0, \;  \forall \; s \in{0,...,D}$) implying that the power demand of the train during the stop is equal to zero, $P_l(t) = 0 \in [t_{l,a}(s),t_{l,d}(s)], \; \forall \; s \in {0,...,D}$.

For each ACC, the operator of train $l$ faces the price of energy given by the set of functions $\lambda_l(t,n) \;  \forall \; t \in [t_{l,n-1},t_{l,n}], \; \forall \; n \in \{1,...,N\}$. The price of energy for train $l$ is therefore composed of the profiles at each ACC:
\begin{align}\label{eq:TrainEnergyPriceProfile}
\lambda_l(t)=\begin{cases} 
\lambda_l(t,1) & t\in[t_{l,0},t_{l,1}]\\
&\vdots\\
\lambda_l(t,N) & t\in[t_{l,N-1},t_{l,N}].
\end{cases}
\end{align}

Note that these profiles are a function of time as well as the $ACC$ traversed by the train. The price faced by the operator of the train can be determined from the Railway Dispatch problem equilibrium prices $\lambda_{n}^{**}$ as follows. Suppose we consider the time horizon $[t_{l,n-1},t_{l,n}]$ over which train $l$ traverses $ACC_n$, and we chose an arbitrary time $t_K \in [t_{l,n-1},t_{l,n}]$, and suppose it corresponds to the $K$'th interval (for example, the time horizon corresponds to $K=1$ in the example shown in Fig. \ref{fig:rDMMtimeline}). Then $\lambda_l(t,n)$ corresponds to the $K$th element of $\lambda_{n}^{e**} \in \R^M$. A similar procedure can be utilized to determine $\lambda_l(t)$ for all intervals in (\ref{eq:TrainEnergyPriceProfile}). In practice, trains may need to preemptively terminate the portion of the schedule included in the Train Dispatch problem or use proxy pricing if they operate within a particular $ACC_n$ outside of the dispatch horizon included in the Asset Dispatch problem. 

The energy cost minimization of the train can now be formulated as: 
\begin{align}
& \underset{x_l,\dot{x_l}}{\text{min}}
&&\int_{\tau=t_0}^{\tau=t_f}P_l(\tau) \lambda_l(\tau) d\tau  \label{eq:TrainObjectiveCostMin}\\
& \text{s.t.} 
&&m_l \ddot{x_l} + F_{DF,l}(\dot{x_l})+\nonumber\\
&&&	\quad m_lgsin(\alpha_l(x_l)) = F_{T,l}\left(\frac{P_l}{ \dot{x_l}}\right)\label{eq:TrainConstraintDynamics}\\
&&&\underline{P_{l}}(x_l) \leq P_l \leq \overline{P_{l}}(x_l) \label{eq:TrainConstraintPower}\\
&&&\underline{F_{T,l}}(\dot{x_l}) \leq F_{T,l} \leq \overline{F_{T,l}}(\dot{x_l}) \label{eq:TrainConstraintTractionForce}\\
&&&\underline{a_{l}} \leq \ddot{x_l} \leq \overline{a_{l}} \label{eq:TrainConstraintAcceleration}\\
&&&\underline{v_{l}}(x_l) \leq \dot{x_l} \leq \overline{v_{l}}(x_l) \label{eq:TrainConstraintVelocity}\\
&&&t_{l,a}(s) \geq \underline{t_{l}}(s), & s \in \{0, ... , D\} \label{eq:TrainConstraintMinArrival}\\
&&&t_{l,d}(s) \leq \overline{t_{l}}(s), & s \in \{0, ... , D\} \label{eq:TrainConstraintMaxDeparture}.	
\end{align}
In the above, (\ref{eq:TrainObjectiveCostMin}) represents the energy cost incurred by the train across all ACCs between $t_0$ and $t_n$. Constraint (\ref{eq:TrainConstraintDynamics}) enforces the train motion dynamics described in Section \ref{subsec:TrainDispatch-DynamicModel}, rearranging (\ref{eq:TrainDynamics}) to solve for the traction force $F_{T,l}$ as a function of the position, velocity and acceleration of the train. The power demand of train $l$ from the traction system at each ACC, $P_l(t)$, is constrained by (\ref{eq:TrainConstraintPower}) which imposes a lower bound at $\underline{P_{l}}(x_l)$ and an upper bound at $\overline{P_{l}}(x_l)$. Note that the limits are functions of the ACC in question as the particular track segment might not be able to receive or provide more than a given power magnitude.

Next, we define the limits of the traction force $F_{T,l}$ in (\ref{eq:TrainConstraintTractionForce}) reducing the feasible traction force window $[\underline{F_{T,l}}(\dot{x_l}),\overline{F_{T,l}}(\dot{x_l})]$ based on the traction force curve of the manufacturer. The acceleration rate of the train $\ddot{x_l}$ is limited per (\ref{eq:TrainConstraintAcceleration}) due to safety considerations of the passengers, who may be standing during moments of deceleration $\underline{a_{l}}$ or acceleration $\overline{a_{l}}$. Similarly, the velocity of the train $\dot{x_l}$ is bound from below by $\underline{v_{l}}(x_l)$ and from above by $\overline{v_{l}}(x_l) $ through equation (\ref{eq:TrainConstraintVelocity}) where the limits are functions of the position $x_l$. This dependency traces back to civil speed limit restrictions which reduce the window of allowed speeds in sections of the track with rail crossings, densely populated areas and passenger stations. Finally, constraints (\ref{eq:TrainConstraintMinArrival}) and (\ref{eq:TrainConstraintMaxDeparture}) represent the schedule constraint of train $l$, achieved by limiting the time spent at the stations to the minimum arrival time $\underline{t_{l}}(s) \forall s \in \{0, ... , f\}$ and the maximum departure time $\overline{t_{l}}(s) \forall s \in \{0, ... , f\}$ respectively.

In summary, in this section we have posed the problem of optimizing the power consumption of a train for a time-schedule in the form of a non-convex constrained optimization problem in (\ref{eq:TrainObjectiveCostMin})-(\ref{eq:TrainConstraintMaxDeparture}). The resulting solution is in the form of power demand profiles $P_l^*(t)$ for a train $l$ at time $t$ for a given set of price profiles $\lambda_l(t)$, constructed by determining the corresponding time interval $K$ and the corresponding equilibrium price $\lambda_{n}^{e**}$. The solution of the Train Dispatch problem may be determined using any one of several commercial software packages such as Matlab's fmincon \cite{matlab2014optimization}. Note that the decision variables used by the optimization problem are  position $x_l$ and velocity $\dot{x_l}$, the two state variables used to represent the dynamical system. Also, numerical solvers can be used to transform the continuous variables (including the decision variables)to discrete-time variables of length $(t_f-t_0)\delta\tau$, where $\delta\tau$ is the time step. 

\section{Overall Transactive Control Architecture - Railway Dynamic Market Mechanism (rDMM)}
\label{sec:TransArch}
In Section II, we posed a combined optimization problem of Railway Dispatch and Train Dispatch in the form of (\ref{eq:ProblemObjective})-(\ref{eq:ProblemConstraintSchedule}). In order to make the problem more tractable, we divided it into two steps, which were addressed in Sections \ref{sec:AgentDispatch} and \ref{sec:TrainDispatch}. In Section \ref{sec:AgentDispatch}, we presented a constrained optimization problem that solves for the optimal dispatch of energy assets, where loads including those of trains as well as renewable generation were assumed to be fixed and the electric and thermal schedules and prices for the generators along the track are determined as $[g_i^{e}(y_{i}^{**}),g_i^{th}(y_{i}^{**})] \in \R^M$ and $\lambda_{n}^{**} = \left[\lambda^{e**}_{n},\lambda^{th**}_{n}\right] \in \R^M$ respectively for dispatch intervals $K=\{1,...M\}$ respectively in (\ref{eq:AgentDispatchObjectiveCostMinimization})-(\ref{eq:AgentDispatchConstraintCapacityLowerBound}). This formulation captures the agent operational cost in the first term of (\ref{eq:ProblemObjective}) and ensures that the agent constraints (\ref{eq:ProblemConstraintElectricPowerBalance})-(\ref{eq:ProblemConstraintAgentCapacityLimits}) are met. In Section \ref{sec:TrainDispatch}, we focused on the optimization of the train power consumption profiles themselves. In particular, we showed how each train operator can solve a constrained optimization problem in (\ref{eq:TrainObjectiveCostMin})-(\ref{eq:TrainConstraintMaxDeparture}) to minimize energy costs given the track prices and required schedule. Minimizing the sum of trains' energy costs is equivalent to the second term of (\ref{eq:ProblemObjective}), subject to the constraints (\ref{eq:ProblemConstraintDynamics})-(\ref{eq:ProblemConstraintSchedule}) for each train. 

\begin{figure}
	\centering
	\includegraphics[width=3.5in]{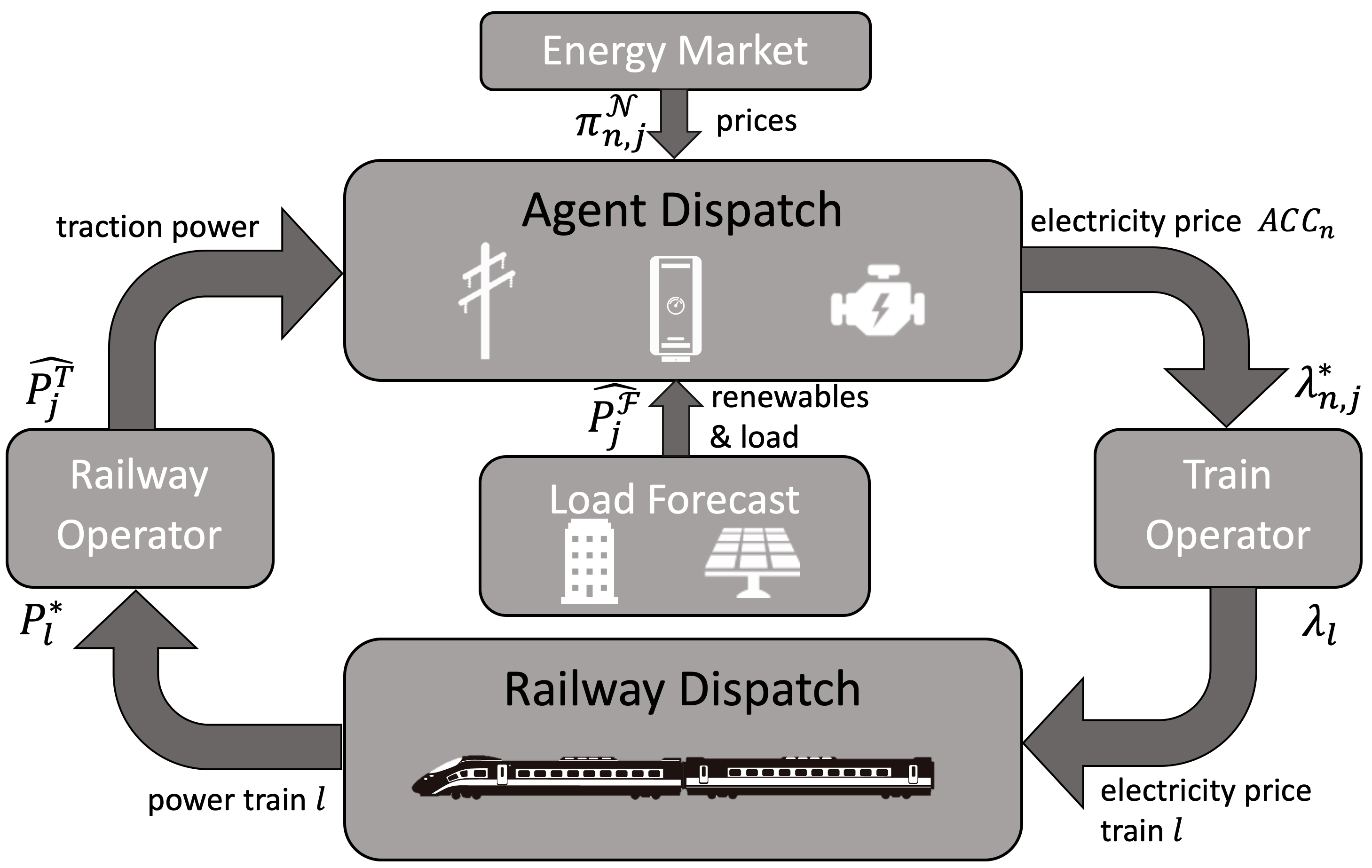}
	\caption{Process flow of the two step $rDMM$ optimization including Railway Dispatch and Train Dispatch. The train and Railway Operators are shown as the intermediaries between the problems that are being solved. Note that train operators and Railway Operators interact in a similar fashion with a growing number of $ACC_n \forall n \in \{1,...,N\}$ and trains $l \in \{1,...,L\}$.}
	\label{fig:rDMMdiagram}
\end{figure}

In this section, we show how the results of Sections \ref{sec:AgentDispatch} and \ref{sec:TrainDispatch} can be interleaved to solve the combined optimization problem posed initially in Section \ref{sec:ProblemFormulation}. This is accomplished by adjusting the power demand of the trains as a function of the prices from the Railway Dispatch problem, as a form of automated Demand Response at fast time-scale, also referred to as transactive control \cite{schweppe1998spot, huang2010analytics}. This transactive control-based $ACC_n$ price provides an incentive for the electric trains to modify their dispatch by iteratively solving the cost minimization problem (\ref{eq:TrainObjectiveCostMin})-(\ref{eq:TrainConstraintMaxDeparture}) with the updated prices. In what follows, we describe the overall transactive control architecture (see Fig.\ref{fig:rDMMdiagram} for an overall schematic and Algorithm \ref{algo:rDMM} for details).

The top block in Fig. \ref{fig:rDMMdiagram} denotes the $Agent$ $Dispatch$, and corresponds to the following functionality: at each forecast instance $j = \{1,...,j^{**}\}$, the $Agent$ $Dispatch$ block takes as inputs the traction power demand $\hat{P}_{j}^T$, passive agent output curves $\hat{P}^\mathcal{F}_j$ and the energy prices from the low voltage side network connections $\pi^\mathcal{N}_{n,j}$, and returns the price profiles $\lambda_{n,j}^*$ and the Railway Dispatch profiles $y_{i,j}^*$. As mentioned in Section \ref{sec:AgentDispatch}, the energy prices from the low voltage side network connections are used to determine the quadratic cost curve of these agents as in (\ref{eq:AgentCostFunction}). Next, these cost curves are used to update the objective function (\ref{eq:AgentDispatchObjectiveCostMinimization}) and the traction and passive agent output curves are used to update constraints (\ref{eq:AgentDispatchConstraintElectricBalance})-(\ref{eq:AgentDispatchConstraintThermalBalance}). The iterative dynamics in (\ref{eq:AgentDispatchObjectiveCostMinimization})-(\ref{eq:AgentDispatchConstraintCapacityLowerBound}) are solved using (\ref{eq:AgentOfferUpdate})-(\ref{eq:AgentPenaltyFunctionUpdate}) for negotiation instances $k=\{1,...,k^*\}$, stopping when the negotiation exit conditions defined earlier in Section \ref{sec:AgentDispatch} are met at each $ACC_n$. 

The block on the right in Fig. \ref{fig:rDMMdiagram} corresponds to the functionality of the $Train$ $Operator$, who composes the price of energy profile for each train $l={1,...,L}$ in (\ref{eq:TrainEnergyPriceProfile}) using the electricity portion of the price profiles from the $Agent$ $Dispatch$ block, by determining the corresponding dispatch interval $K$ and the corresponding element of the price vector $\lambda_{n}^{e**}$. These are then used in the $Train$ $Dispatch$ block (bottom) to determine the next dispatch profile forecast for each train $P_l^*$.

The bottom block in Fig. \ref{fig:rDMMdiagram} denotes the $Train$ $Dispatch$ and uses the prices $\lambda_l(t)$ as in (\ref{eq:TrainEnergyPriceProfile}) to determine the cost-minimizing train dispatch for every train. This is accomplished using the optimization procedure discussed in Section \ref{sec:TrainDispatch}: the energy price profiles from the $Agent$ $Dispatch$ block are used to update the objective function of the Train Dispatch problem in (\ref{eq:TrainObjectiveCostMin}). Once this update is complete, a new power demand profile of the train $P_l^*$ is determined by solving (\ref{eq:TrainObjectiveCostMin})-(\ref{eq:TrainConstraintMaxDeparture}). 

The block on the left in Fig. \ref{fig:rDMMdiagram} represents the $Railway$ $Operator$, who collects the new power demand profiles for each of the trains $l={1,...,L}$ after each forecast instance $j$ and assembles the total traction power demand for each $ACC$ in (\ref{eq:AgentTractivePower}). This new profile is used in the negotiations (\ref{eq:AgentOfferUpdate})-(\ref{eq:AgentPenaltyFunctionUpdate}), in addition to new forecasts that may become available from other passive agents $\hat{P}^{F}_{j} = [\hat{P}^{re}_{j}+\hat{P}^e_{j},\hat{P}^{th}_{j}]$ at each $j$ and energy prices from the low voltage side network connections $\pi^\mathcal{N}_{n,j}$ used to determine the cost function parameters $a_{i,K}$, $b_{i,K}$, and $c_{i,K}$ in (\ref{eq:AgentCostFunction}) for $i\in\mathcal{N}$.

The cycling between the $Agent$ $Dispatch$ and $Train$ $Dispatch$ blocks repeats for forecast instances $j = \{1,...,j^{**}\}$. If the resulting price $\lambda_{n,j}^*$ and Railway Dispatch $y_{i,j}^*$ profiles meet the forecast exit conditions for the network ($\left|y_{i,j+1}^*-y_{i,j}^*\right|\geq\gamma_j^y \; \& \; \left|\lambda_{n,j+1}^*-\lambda_{n,j}^*\right|\geq\gamma_j^\lambda$), then the algorithm stops, dispatching the agents at the last negotiation equilibrium $y_i^{**}$ and are compensated based on the price profiles $\lambda_{n}^{**}$ from the top block. Similarly, the trains are dispatched based on the last forecast update $P^{T**}$ from the bottom block and their operators are required to pay the last negotiation equilibrium price $\lambda_{n}^{**}$. As mentioned before, the overall iteration is ensured to stop by using suitable exit conditions. A settlement procedure may be designed to collect the payments of the agents at a slower frequency than the convergence of the $rDMM$ (i.e. monthly payments). Once dispatch takes place for $K=1$ the dispatch horizon shifts by one interval, and the procedure set forth with $j=1$ starts again. 

\begin{algorithm}
	\begin{algorithmic}
		\STATE$j=1$; $\lambda_{n,0}=0$\\
		\WHILE{$\left|y_{i,j}^*-y_{i,j-1}^*\right|>\gamma_j^y \; OR \; \left|\lambda_{n,j}^*-\lambda_{n,j-1}^*\right|>\gamma_j^\lambda$}
		\FOR{$l=[1,...,L]$} 
		\STATE Update (\ref{eq:TrainEnergyPriceProfile}) using $\lambda_{n,j}^*$\\
		\STATE Solve (\ref{eq:TrainObjectiveCostMin})-(\ref{eq:TrainConstraintMaxDeparture}) for $P_l^*$\\
		\ENDFOR
		\FOR{$n=[1,...,N]$}
		\STATE Update (\ref{eq:AgentTractivePower}) using $P_l^* \;\forall\; l \;\in \;[1,...,L]$
		\STATE Update (\ref{eq:AgentDispatchObjectiveCostMinimization})-(\ref{eq:AgentDispatchConstraintThermalBalance}) using $\hat{P}^T_j$, $\hat{P}^\mathcal{F}_j$, $\pi_{n,j}^\mathcal{N}$\\
		\STATE$k=0$\\
		\WHILE{$\left|\lambda_{n,k+1}-\lambda_{n,k}\right|>\gamma_k$}
		\STATE Solve for $y_i^{k+1}$ in (\ref{eq:AgentOfferUpdate}) \\
		\STATE Solve for $\lambda^e_{n,k+1}$ in (\ref{eq:AgentElectricPriceUpdate}) and $\lambda^{th}_{n,k+1}$ in (\ref{eq:AgentThermalPriceUpdate})\\
		\STATE $k++$\\
		\ENDWHILE
		\STATE Update $\lambda_{n,j}=\lambda_{n,j+1}$ and $\lambda_{n,j+1}=\lambda_{n,k^*}$
		\ENDFOR 
		\STATE $j++$\\
		\ENDWHILE
		\STATE Dispatch $\lambda_{n}^{**}$, $y_{i}^{**}$,  $P^{T**}$ for $K=1$.
	\end{algorithmic}
	\caption{Railway Dynamic Market Mechanism $rDMM$}\label{algo:rDMM}
\end{algorithm}

\section{Simulations}
\label{sec:Simulations}

\begin{figure}
	\centering
	\includegraphics[width=3.5in]{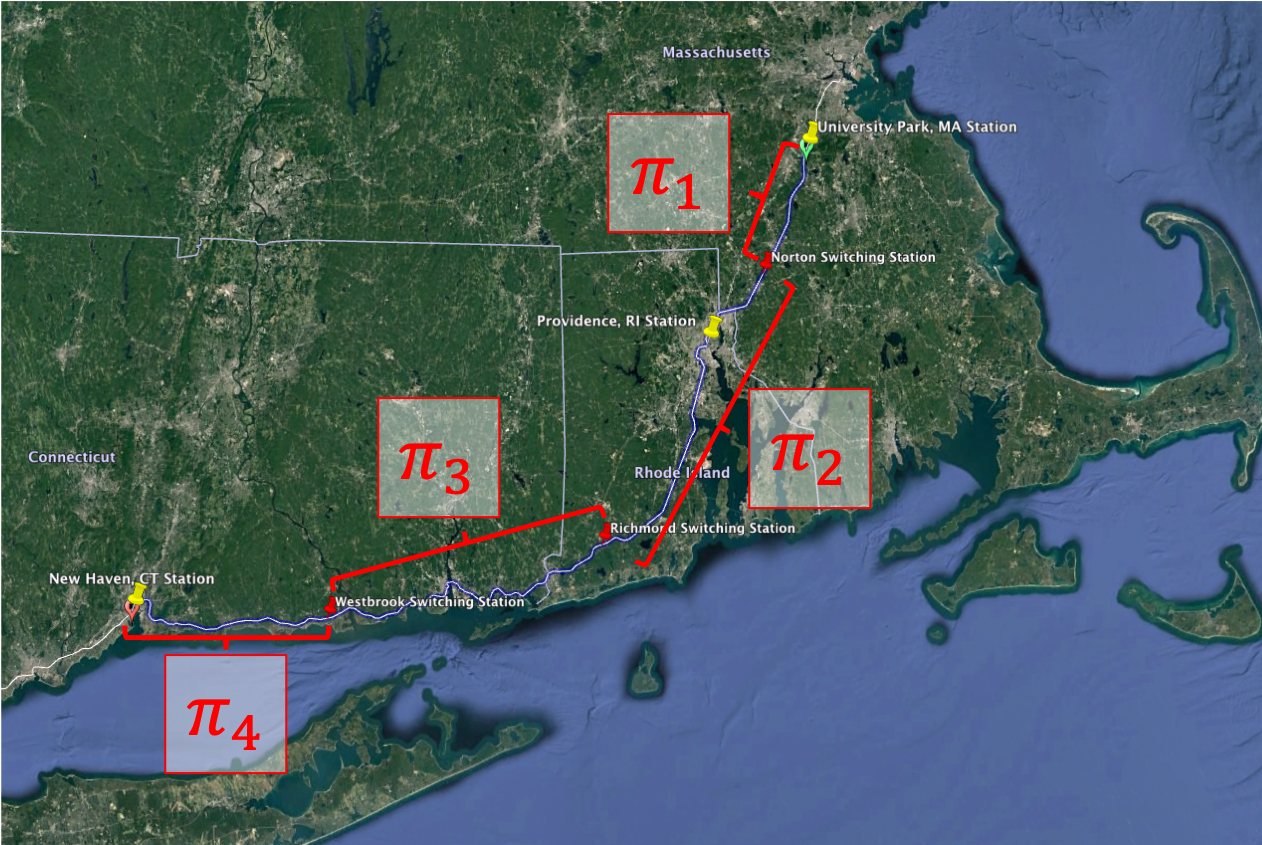}
	\caption{Map of the four pricing regions identified along Amtrak's NEC between University Park Station in Massachusetts and New Haven Station in Connecticut. This graphic was developed using Google Earth Pro \cite{googleEarth}.}
	\label{fig:PricingRegionsTracks}
\end{figure}

The northern Amtrak NEC between Boston, MA and New Haven, CT (within the ISO-NE power system) emerges from a review of the electric railway systems in the United States as a prime case study for our analysis, due to its four segmented rail power zones that result in the pricing regions identified in Fig. \ref{fig:PricingRegionsTracks}. The four area control centers, $ACC_n$ identified with $n\in\{1,2,3,4\}$ are powered by the substations at Sharon, MA; New Warwick, RI; London, CT; and Branford, CT respectively and are considered separate pricing regions, each with price $\pi_n \forall n \in \{1,2,3,4\}$. 

\begin{figure}
	\centering
	\includegraphics[width=3.5in]{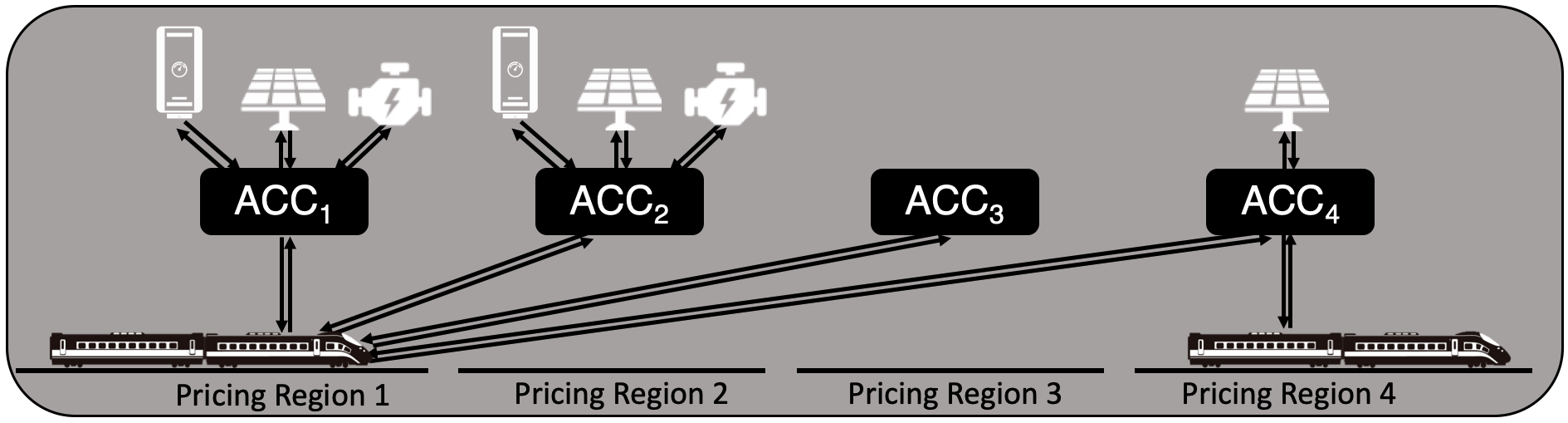}
	\caption{Schematic of the Area Control Centers proposed for Amtrak's northend NEC used in simulation. The four $ACCs$ have varying levels of load, renewable deployment, dispatchable agents and network energy pricing.}
	\label{fig:ACCschematic}
\end{figure}

Using the location of these substations, publicly available electric utility tariff information \cite{eversourcema2018tariff,nationalgrid2018tariff,eversourcect2018tariff} and real-time energy market data from ISO-NE \cite{isoneLMP}, we estimate the real-time energy cost to the operator at each one of the four $ACC_n$ which are in turn used as the low-voltage side network connection costs $\pi^\mathcal{N}_{n,j}$ of the four network connection agents $\mathcal{N}_n$ that are used to update the quadratic cost curve of these agents in (\ref{eq:AgentCostCurve}). Based on the characteristics of the Route 128/University Park ($ACC_1$) and Providence ($ACC_2$) passenger stations and their surrounding commercial spaces we also add cogeneration assets $\mathcal{C}_n$ and boilers $\mathcal{H}_n$ using NREL's System Advisory Model (SAM) \cite{nrel2017sam} for sizing and estimation of the quadratic cost function coefficients in (\ref{eq:AgentCostCurve}). 

Due to the large roofs and parking lots near the University Park, Providence and New Haven passenger stations, we also assumed that three PV solar arrays ${re}_n$ can be added at $ACC_1$, $ACC_2$ and $ACC_4$, and use SAM alongside satellite imagery to size the arrays and estimate their yearly production. We also used the satellite imagery to measure the footprint of the passenger stations and estimate the electric ${e}_n$ and thermal ${th}_n$ load using the EPA commercial building templates that can be accessed through SAM.

Amtrak's high-speed Acela Express service along the NEC utilizes high-speed locomotives developed by Bombardier in the late 1990s based on the French TGV \cite{bombardier2001acela}. Acela Express trains have a total empty weight of $531.2 MT$ and a full capacity weight of $556.7 MT$. In simulation we assume a partially occupied weight of $m_l = 545 MT$. Although the Acela trains are designed to achieve a $264 km/h$ top speed, they are limited in operation to $240 km/h$ which is equivalent to $v_{l,max} = 66.67 m/s$. 

The maximum train traction power $P_{l,max}$ is $9.2 MW$, while regenerative braking is limited in operation to $P_{l,min}=-6.0MW$ \cite{yu2007measurement}. In the absence of public data on the acceleration rates of high-speed trains like the Acela, the estimates used in our simulation ($a_{l,min}=-0.5 ms^{-2}$ and $a_{l,max}=0.5 ms^{-2}$) were adopted from models of electric train systems used by EPRail. Fitting the Davis equation (\ref{davisEquation}) to the Acela Express drag and rail friction curve we have that $A_l=10,195.16$, $B_l=65.81$, and $C_l=25.02$.

Using the pricing information of the nodes along the track and the Amtrak Acela Express train timetable \cite{amtrak2018timetable}, we simulate a train following a power profile that minimizes total work as a baseline and a train dispatched by the $rDMM$ methodology summarized in Algorithm \ref{algo:rDMM}.

\begin{figure}[tp]
	\centering
	\smallskip
	\includegraphics[width=2.5in]{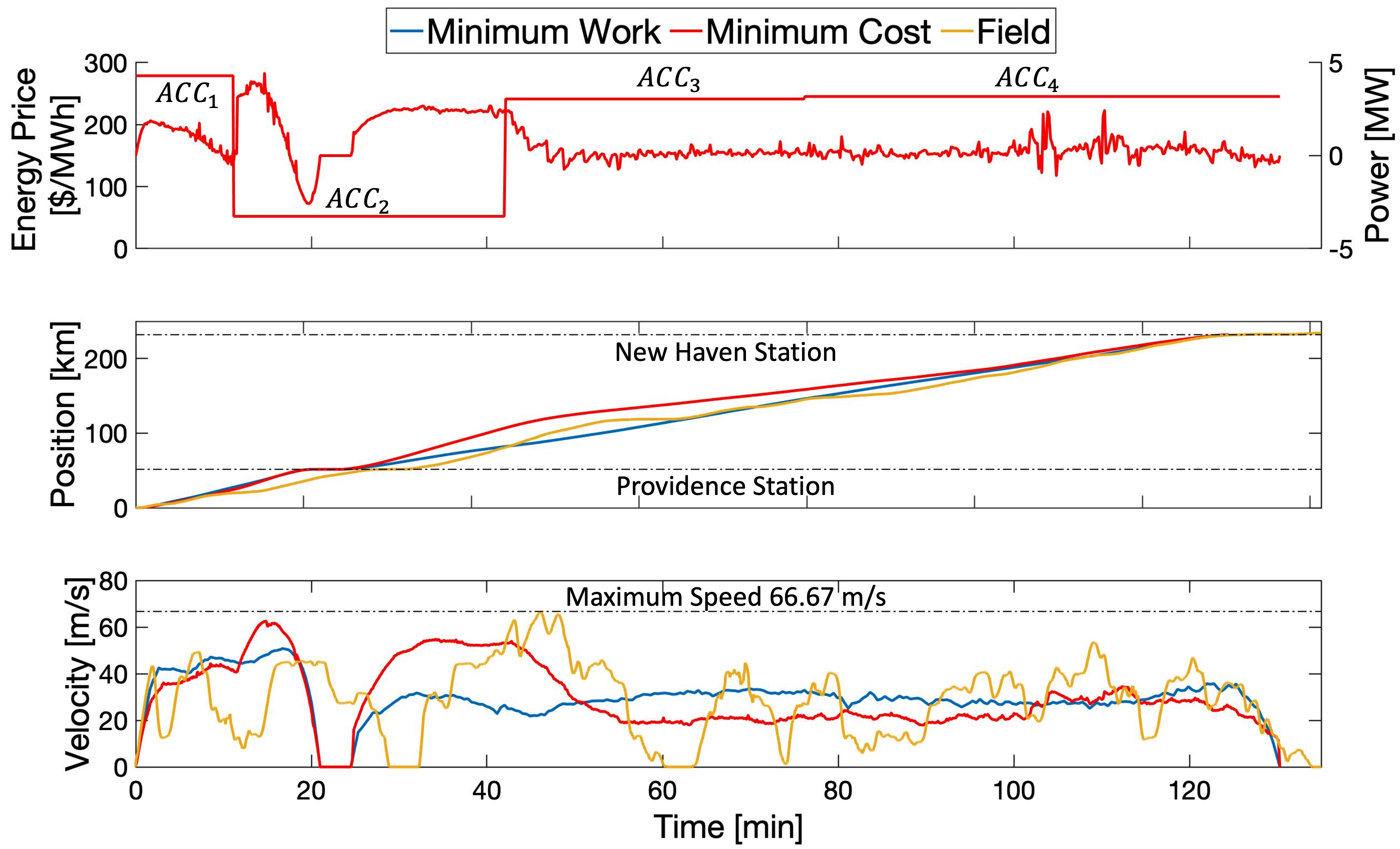}
	\caption{Plot of average energy price  [\$/MWh], position [km] and velocity [m/s] against time for the train that minimizes work (blue), the one dispatched following $rDMM$ (red), and the field train (yellow) of a southbound trip on Amtrak Acela between University Park Station in MA and New Haven Station in CT with a stop in Providence Station in RI. The power trajectory [MW] for the train dispatched by the $rDMM$ methodology is also plotted on the price plot, showcasing the power injection from the train into the electric railway during regenerative braking.}
	\label{WholesaleMarketSimulation}	
\end{figure}

We now report the results obtained using the rDMM outlined in Section V, which we will denote as the transactive controller. In particular Algorithm \ref{algo:rDMM} was run with all numerical parameters as in Table \ref{table:NumericalParameters}. The results are shown in Fig. \ref{WholesaleMarketSimulation} for a single train’s travel profile corresponding to for the 6:21AM University Park departure of Acela 2155 on January 18, 2018, a day that exhibited large network pricing differentials. These results include the energy prices $\lambda_n$ as the train traverses the four ACCs, and the position and velocity profiles of the train. It can be seen that the train schedules are met, and the velocity limits are accommodated. The most interesting result corresponds to the energy price shown in the top plot in Fig. \ref{WholesaleMarketSimulation}, and corresponds to the minimization of the cost function in Eq. (\ref{eq:AgentDispatchObjectiveCostMinimization}). This price profile in turn leads to an optimized cost in Eq. (\ref{eq:TrainObjectiveCostMin}) of $\$200.62$ for this single train travel.

\begin{center}
	\begin{tabular}{PIIIIU}
		\toprule
		\multicolumn{1}{P}{\textbf{$ACC_n$}} & \multicolumn{1}{I}{Agent} & \multicolumn{1}{I}{\textbf{$\overline{P_i^{e,th}}$}} & \multicolumn{1}{I}{\textbf{$d_i^e$}} & \multicolumn{1}{I}{\textbf{$d_i^{th}$}} & \multicolumn{1}{U}{\textbf{$b_{i,K}$}} 
		\tabularnewline
		\midrule
		\multicolumn{1}{P}{1} & \multicolumn{1}{I}{$\mathcal{H}_1$} & \multicolumn{1}{I}{10,432} & \multicolumn{1}{I}{0} & \multicolumn{1}{I}{1} & \multicolumn{1}{U}{0.0303} 
		\tabularnewline	
		\multicolumn{1}{P}{1} & \multicolumn{1}{I}{$\mathcal{C}_1$} & \multicolumn{1}{I}{1,550} & \multicolumn{1}{I}{1} & \multicolumn{1}{I}{1.02} & \multicolumn{1}{U}{0.0629} 
		\tabularnewline	
		\multicolumn{1}{P}{1} & \multicolumn{1}{I}{$\mathcal{N}_1$} & \multicolumn{1}{I}{10,000} & \multicolumn{1}{I}{1} & \multicolumn{1}{I}{0} & \multicolumn{1}{U}{$\pi_{1,j}$} 
		\tabularnewline		
		\midrule
		\multicolumn{1}{P}{2} & \multicolumn{1}{I}{$\mathcal{H}_2$} & \multicolumn{1}{I}{20,864} & \multicolumn{1}{I}{0} & \multicolumn{1}{I}{1} & \multicolumn{1}{U}{0.0303} 
		\tabularnewline	
		\multicolumn{1}{P}{2} & \multicolumn{1}{I}{$\mathcal{C}_2$} & \multicolumn{1}{I}{4,560} & \multicolumn{1}{I}{1} & \multicolumn{1}{I}{2} & \multicolumn{1}{U}{0.0818} 
		\tabularnewline	
		\multicolumn{1}{P}{2} & \multicolumn{1}{I}{$\mathcal{N}_2$} & \multicolumn{1}{I}{10,000} & \multicolumn{1}{I}{1} & \multicolumn{1}{I}{0} & \multicolumn{1}{U}{$\pi_{2,j}$} 
		\tabularnewline		
		\midrule
		\multicolumn{1}{P}{3} & \multicolumn{1}{I}{$\mathcal{N}_3$} & \multicolumn{1}{I}{10,000} & \multicolumn{1}{I}{1} & \multicolumn{1}{I}{0} & \multicolumn{1}{U}{$\pi_{3,j}$} 
		\tabularnewline		
		\midrule
		\multicolumn{1}{P}{4} & \multicolumn{1}{I}{$\mathcal{N}_4$} & \multicolumn{1}{I}{10,000} & \multicolumn{1}{I}{1} & \multicolumn{1}{I}{0} & \multicolumn{1}{U}{$\pi_{4,j}$} 
		\tabularnewline		
		\bottomrule
	\end{tabular}
	\captionof{table}{
	Agent numerical parameters used in simulation. Maximum capacity is expressed for the binding thermal or electric characteristic and can be identified by the mapping coefficient $d_i^e$ or $d_i^{th}$ that is equal to 1. The other cost function parameters in \ref{eq:AgentCostFunction} were used at a minimum in simulation, setting $c_i$ to a small positic quantity.}
	\label{table:NumericalParameters}
\end{center}

In order to evaluate the optimality of the proposed transactive architecture, we compare the above results with two other profiles which are shown in Fig. \ref{WholesaleMarketSimulation}. The first one corresponds to a field dataset that was
collected using the GPS of a mobile phone and the MyTracks iOS application \cite{myTracks} on the Acela 2171. It can be seen that the maximum speed of $66.7m/s$ as well as the position datasets corresponding to the transactive controller are consistent with our field dataset. We computed the corresponding train cost for this field dataset assuming the same dynamical model for the Acela train employed in simulation, summarized in Equation (\ref{eq:TrainConstraintDynamics}), which allows us to solve for the traction force $F_{T,l}$ as a function of the position, velocity and acceleration datasets collected with the phone GPS and accelerometers. This force dataset can in turn be used in conjunction with the volatility dataset to arrive at the tractive power profile for the route. Finally, the marginal cost along the route, set by the low voltage side network connection cost, can be applied to the profile to determine the total cost. It was observed that this cost was $\$865.01$, which shows that our rDMM results in a $75\%$ reduction. It is possible that the actual reduction from the rDMM may be somewhat smaller, as we have not incorporated other speed limit restrictions along the track such as rail crossings and densely populated areas in the simulation of the rDMM. 

The second profile shown in Fig. \ref{WholesaleMarketSimulation}, denoted as minimum work, was obtained by solving the Train Dispatch problem in (\ref{eq:TrainObjectiveCostMin})-(\ref{eq:TrainConstraintMaxDeparture}) with a uniform price profile $\lambda_l(\tau) = 1$. This yielded a position and velocity profile as shown in Fig. \ref{WholesaleMarketSimulation}. The corresponding price profile is shown in Fig. \ref{WholesaleMarketSimulation} as well, which led to a total trip cost $\$273.69$. Note that the train dispatched by the $rDMM$ methodology can achieve a $25\%$ cost reduction when compared to the train dispatched under the standard minimum work (from $\$273.69$ to $\$205.62$), providing an initial estimate of the value of incorporating dynamic, price-responsive train dispatch in the electric railway operation.

To our knowledge, the current procedure adopted by Amtrak for the train profile does not have such an optimization approach, but rather allows train operators to accelerate and decelerate the train at their discretion with the supervision and intervention of the Positive Train Control system. The approach in \cite{eldredge2011trip}, implemented in freight locomotives, employ algorithms similar to the Minimum Work method reported above. 

\section{Conclusions and Future Works}
\label{sec:Conclusion}
Electric trains are a major untapped source of demand-side flexibility in electricity networks. Our findings contribute to the evolution of transportation control systems devoted to work minimization toward higher-level objectives such as the social welfare maximization of joint transportation-electric infrastructures. In particular, our proposed 2-step optimization of railway dispatch of all DER agents along the train track followed by train dispatch, facilitated by coordinated operations of the Railway Operator and Train Operator suggests that the inclusion of time and space varying pricing information modifies the optimal power profiles of DERs and trains, yielding reductions in electricity costs for relatively small increases in work. Simulation studies of the Southbound Amtrak service along the Northeast Corridor in the United States shows a 25\% reduction in energy costs when compared to standard trip optimization based on minimum work, and 75\% reduction in energy costs when compared to the train cost calculated using a field dataset. 

Fundamentally, the rDMM introduces transactive energy as an additional degree of freedom in the control of a system, capitalizing on technology advancement (e.g. communication cost reductions, GPS, widespread adoption of regenerative braking) to bridge the objectives of individual agents (e.g. trains, DERs) with those of global infrastructure (e.g. traction system, wholesale energy markets). That is, through adjustments of incentives in the form of electricity prices, we were able to ensure a coordinated set of profiles for all DERs and trains.

This technology could further motivate the deployment of automation technologies in train systems, as the business case improves when factoring electrical cost reductions. We expect that our findings could be developed into a software package for train operators, similar to GE’s Trip Optimizer technology which has been adopted by heavy haul train operators to decrease fuel use \cite{eldredge2011trip}. 

\subsubsection{Demand Charge Management}
Although our work is a step towards including the electric traction system's costs within the Train Dispatch problem, we only reflect energy-related costs $(\$/MWh)$. In reality the Railway Operator will also incur demand or capacity charges $(\$/MW)$ from the utility or ISO. These charges can also be reduced, in principle, using a transactive control methodology where the incentive signal shifts and smooths the power profile of the individual trains such that a reduction in demand at the main interconnection (traction substation) is met. We considered including the demand charge component within the cost function of low voltage side network connections $\mathcal{N}_n$ as a second order term, but this methodology did not appropriately capture the time-scale (typically months) at which demand charges are evaluated. A means of achieving demand charge management is to update constraints within the optimization problem, modifying the minimum and maximum power limits $\underline{P_{l}}(x_l)$ and $\overline{P_{l}}(x_l)$ in (\ref{eq:TrainConstraintPower}).

\subsubsection{Regulation and Reserve Market Participation}
Similar to the research direction regarding demand charges, we would like to extend the services provided by the transactive control system to the electrical network beyond energy and onto products for frequency regulation and operational reserves. Practical limitations of providing these services as well as the incentive and compensation mechanisms remain to be explored.

\subsubsection{Mass Transit Systems}
Our analysis focused on the practical intricacies of high-speed rail systems. Although our simulations were based on the high-speed rail example in the United States, the Amtrak Acela service, we are aware that other systems, such as the MBTA mass transit T service also evidences discontinuities in  energy price along their track and have strategic plans to add rail-side generation \cite{donaghy2015mbta}. Extending our simulation work to mass transit systems would widen the applicability of our proposed transactive control architecture. 

\bibliographystyle{IEEEtran}
\bibliography{20200614_arXiv_rDMM}

\clearpage
\begin{IEEEbiography}[{\includegraphics[width=1in,height=1.25in,clip,keepaspectratio]{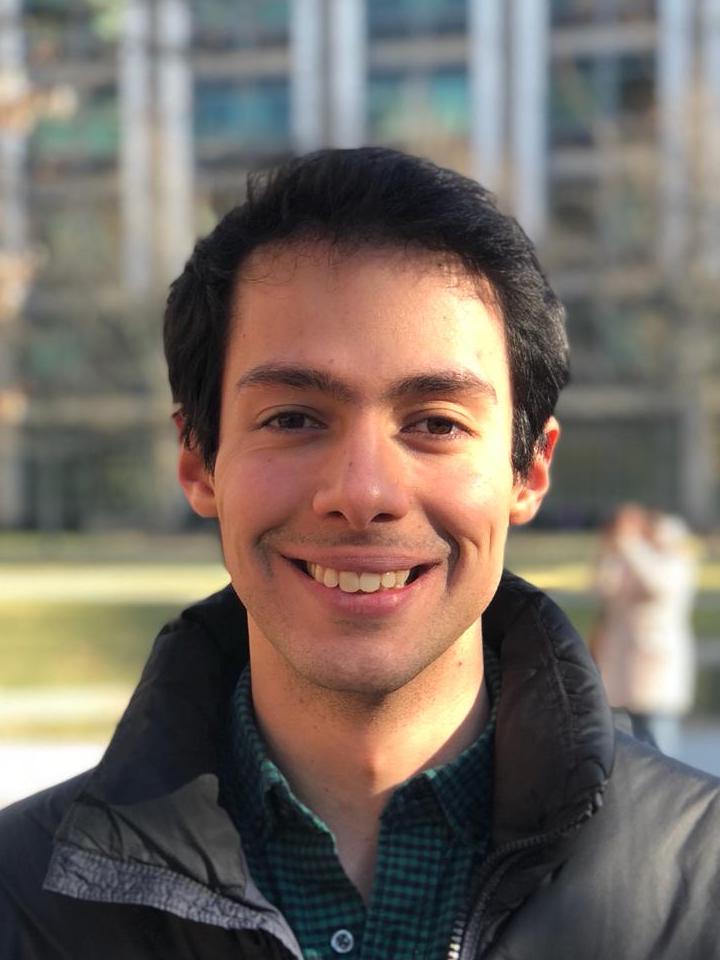}}]{David D'Achiardi} is a Research Assistant at the Active-Adaptive Control Laboratory in the Department of Mechanical Engineering at the Massachusetts Institute of Technology (MIT). He received his M.S. in Mechanical Engineering in 2019 from MIT. Prior to that, he worked as a Mechanical Design Engineer at Tesla’s Gigafactory in Reno, Nevada. He received his B.S. in Mechanical Engineering and Economics in 2016 from MIT. David is a recipient of the 2017 Douglas and Sara Bailey Scholarship and the 2015 Carl G. Sontheimer Prize for Excellence in Innovation and Creativity in Design from the Department of Mechanical Engineering at MIT. His research interests include energy market design under high renewable power adoption and transactive control of electric railway systems with distributed energy resources.
\end{IEEEbiography}

\begin{IEEEbiography}[{\includegraphics[width=1in,height=1.25in,clip,keepaspectratio]{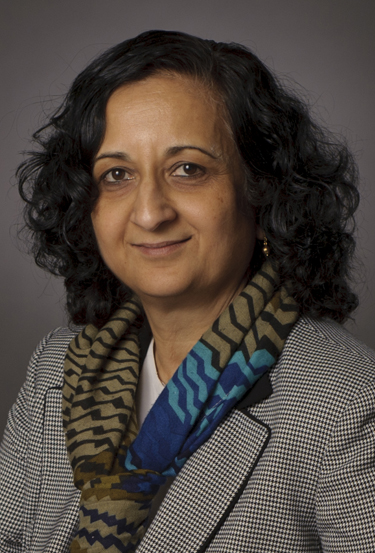}}]{Anuradha M. Annaswamy}
	is Founder and Director of the Active-Adaptive Control Laboratory in the Department of Mechanical Engineering at MIT.  Her research interests span  adaptive control theory and its applications to aerospace, automotive, and propulsion systems as well as cyber physical systems such as Smart Grids, Smart Cities, and Smart Infrastructures. Dr. Annaswamy is an author of over 100 journal publications and 250 conference publications, co-author of a graduate textbook on adaptive control,  and co-editor of several cutting edge science and technology reports including \textit{Systems \& Control for the future of humanity, research agenda: Current and future roles, impact and grand challenges} (Annual Reviews in Control, 2016), \textit{Smart Grid Control: Overview and Research Opportunities} (Springer, 2018), and \textit{Impact of Control Technology} (IoCT-report 2011 and 2013). 
	Dr. Annaswamy has received several awards including the George Axelby (1986) and Control Systems Magazine (2010) best paper awards from the IEEE Control Systems Society (CSS), the Presidential Young Investigator award from NSF (1992), the Hans Fisher Senior Fellowship from the Institute for Advanced Study at the Technische Universität München (2008), the Donald Groen Julius Prize from the Institute of Mechanical Engineers (2008). Dr. Annaswamy has been elected to be a Fellow of the IEEE (2002) and IFAC (2017). She received a Distinguished Member Award and a Distinguished Lecturer Award from IEEE CSS in 2017. 
	Dr. Annaswamy is actively involved in IEEE, IEEE CSS, and IFAC. She has served as General Chair of the American Control Conference (2008) as well as the 2nd IFAC Conference on Cyber-Physical \& Human Systems (2018). She is Deputy Editor of the Elsevier publication Annual Reviews in Control (2016-present). She has been a member of the IEEE Fellows Committee and the IEEE CSS Outreach Committee, and is the Chair of IEEE Smart Grid Meetings and Conferences. In IEEE CSS, she has served as Vice President of Conference Activities (2015-16) and Technical Activities (2017-18), and will serve as the President in 2020. 
\end{IEEEbiography}

\begin{IEEEbiography}[{\includegraphics[width=1in,height=1.25in,clip,keepaspectratio]{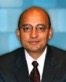}}
	]{Sudip K. Mazumder} received the Ph.D. degree in electrical and computer engineering from Virginia Tech, Blacksburg, VA, USA, in 2001 and M.S. degree in electrical power engineering from Rensselaer Polytechnic Institute, Troy, NY, USA, in 1993. He is a Professor and Director of the Laboratory for Energy and Switching-Electronic Systems at the University of Illinois at Chicago (UIC), USA, since 2001 and is the President of NextWatt LLC since 2008. He has about 30 years of professional experience and has held R\&D and design positions in leading industrial organizations and has served as a Technical Consultant for several industries. His current areas of research interests are advanced control of power/energy systems and wide-bandgap power electronics and devices.

	He has published more than 225 refereed papers, delivered 100 keynote/plenary/distinguished/invited presentations, and received and carried out over 50 sponsored research since joining UIC. At the University of Illinois, he is the recipient of Distinguished Researcher of the Year Award (2020), Inventor of the Year Award (2014), and University Scholar Award – highest award of the university (2013). He also received the ONR YIP and NSF CAREER Awards (2005, 2003), and 5 IEEE Awards. 

	In 2016, he was elevated to the rank of an IEEE Fellow. He also served as a Distinguished Lecturer for IEEE Power Electronics Society (PELS) between 2016-2019. Since 2019, he is also the EiC-at-Large for IEEE Transactions on Power Electronics, the leading journal in power electronics in the world. Currently, he is a Member-at-Large and AdCom Member for IEEE PELS and serves as the Chair for IEEE PELS Technical Committee on Sustainable Energy Systems as well. He is the Chair for 2021 IEEE Power Electronics for Distributed Generation Conference (PEDG’21).

\end{IEEEbiography}

\begin{IEEEbiography}[{\includegraphics[width=1in,height=1.25in,clip,keepaspectratio]{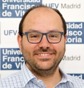}}]{Eduardo Pilo} 
received his PhD degree in ICAI (Engineering School of the Pont. Univ. of Comillas, Madrid, Spain) in 2003. From 2003 to 2010, he served as a researcher in the Institute for Research in Technology (Comillas), where he was in charge of all the research projects related with railway power systems. From 2010 to 2012, he worked for the electrical industry as a consultant of the company Multitest09.  In 2012, he founded his own company, EPRail, focused on providing research and consultancy services in the field of power systems and railways. From Jun 2013 to Jul 2014, he also served in the Univ. of Illinois at Chicago as a Visiting Professor and Research Scientist. Starting in May 2019, he currently serves as a full-time professor in the Universidad Francisco de Vitoria (Madrid, Spain).

His expertise comprises the modelling, simulation and analysis of railway power systems, including their interrelations with transmission \& distribution grids. He has conducted research and consultancy for major companies in the railway sector (such as ADIF, Renfe, ACS, Red Eléctrica de España, Bombardier, etc.) and public bodies (Ministry of Public Works, European Commission, etc.). He has co-authored 40+ journal and conference articles and book chapters in this field. In 2010, he was awarded with the 10th Talgo Prize for Technological Innovation.

\end{IEEEbiography}

\end{document}